\documentclass[aps,pra,twocolumn,amsmath,amssymb,floatfix,longbibliography]{revtex4-2}

\usepackage{amsmath}
\usepackage{txfonts}
\usepackage{microtype}
\usepackage{graphicx}
\usepackage{color}
\usepackage{ulem}
\usepackage{bm}
\usepackage[english]{babel}
\usepackage{braket}
\usepackage{mathtools,nicefrac}
\usepackage[caption=false]{subfig}

\usepackage{amssymb}
\usepackage{epstopdf}
\usepackage{braket}
\usepackage{bbm}
\usepackage{subfig}

\usepackage{cancel}

\def\be{\begin{equation}}
\def\ee{\end{equation}}
\def\bea{\begin{eqnarray}}
\def\eea{\end{eqnarray}}

\def\br{\ensuremath{\mathbf{r}}}

\def\bE{\ensuremath{\mathbf{E}}}

\def\bA{\ensuremath{\mathbf{A}}}
\def\bP{\ensuremath{\mathbf{P}}}
\def\bp{\ensuremath{\mathbf{p}}}
\def\bk{\ensuremath{\mathbf{k}}}

\def\au{\ensuremath{\textnormal{ a.u.}}}

\def\intinfty{\ensuremath{\int^{\infty}_{-\infty}}}

\newcommand*\diff{\mathop{}\!\mathrm{d}}

\DeclareMathAlphabet\mathbfcal{OMS}{cmsy}{b}{n}

\begin{document}

\title{Sub-barrier recollisions and  the three classes of tunneling time delays in strong-field ionization}

\author{Michael Klaiber}
\email{klaiber@mpi-hd.mpg.de}
\author{Daniel Bakucz Can\'{a}rio}
\author{Karen Z. Hatsagortsyan}
\email{k.hatsagortsyan@mpi-hd.mpg.de}
\affiliation{Max-Planck-Institut f{\"u}r Kernphysik, Saupfercheckweg 1, 69117 Heidelberg, Germany}

\date{\today}

\def\be{\begin{equation}}
\def\ee{\end{equation}}
\def\bea{\begin{eqnarray}}
\def\eea{\end{eqnarray}}
\begin{abstract}
Tunneling ionization is characterized by a negative time delay, observed asymptotically as a specific shift of the photoelectron momentum distribution,
%This shift corresponds to a negative time delay
which is caused by the interference of the sub-barrier recolliding and direct ionization paths.
%, as laid out in [Phys. Rev. Lett. 120, 013201 (2018)].
In contrast, a \textit{Gedankenexperiment} following the peak of the wavefunction shows a positive tunneling time delay at the tunnel exit, considering only the direct ionization path. In this paper we investigate the effects of sub-barrier recollisions on the time delay pattern at the tunnel exit. We conclude that the interference of the direct and recolliding trajectories  decreases the tunneling time delay at the exit by the value  equal to the asymptotic time delay maintaining, however, its sizeable positive value.
% Nevertheless, the tunneling time delay at the exit  maintains its sizeable positive value.
Finally, we discuss the recent experiment [Light: Science \& Applications 11, 1 (2022)] addressing the tunneling time in a modified two-color attoclock setup. The analysis of the experimental findings with our theoretical model indicates the physical necessity to introduce a new  time characteristic for tunneling ionization -- the time delay describing the  initiation of the tunneling wave packet.

\end{abstract}

\maketitle

\section{Introduction}

The tunneling ionization is at the heart of attoscience \cite{Corkum_2007,Krausz_2009}. The state-of-the-art attoclock technique \cite{Maharjan_2005,Eckle_2008a} provides an exceptional time resolution of the order of tens of attoseconds via mapping the time to the attoclock offset angle in the angular streaking process. This allows for an experimental investigation of the time delay problem during the quantum tunneling process \cite{Eckle_2008b, Pfeiffer_2012, Landsman_2014o, Sainadh_2019,Camus_2017}. This fundamental question raised an extensive discussion \cite{Eckle_2008b, Pfeiffer_2012, Landsman_2014o, Sainadh_2019, Camus_2017, Teeny_2016a, Teeny_2016b, Yakaboylu_2013, Han_2019, Orlando_2014, Orlando_2014PRA, Lein_2011, Landsman_2014, Torlina_2015, Ni_2018b, Ni_2018a,Ni_2016,Hofmann_2021,Landsman_2015,Zheltikov_2016,Zimmermann_2016, Liu_2017, Song_2017,Yuan_2017, Klaiber_2018, Eicke_2018, Douguet_2018, Bray_2018, Crowe_2018, Ren_2018, Tan_2018, Sokolovski_2018, Quan_2019, Douguet_2019, Hofmann_2019, Serov_2019, Wang_2019, Yuan_2019, Kheifets_2020, Kolesik_2020,Hofmann_2021,Yu_2022,Klaiber_2022}, which can be more appreciated in the context of the general problem of the tunneling time \cite{MacColl_1932,Eisenbud,EisenbudWigner,wigner_1955,Smith_1960,Baz_1966,
 Butttiker_1982,Pollak_1984,Steinberg_1995,Czirjak_2000,Peres_1980,Landauer_1994,Hauge_1998,
 Muga,deCarvalho,Davies_2005}, with different tunneling time definitions corresponding to different type of measurements: Eisenbud-Wigner \cite{Eisenbud,EisenbudWigner,wigner_1955,Smith_1960}, B\"uttiker–Landauer \cite{Butttiker_1982}, Pollak-Miller \cite{Pollak_1984}, and Larmor \cite{Baz_1966} tunneling times, the latter being
 recently measured for cold atoms  \cite{Ramos_2020}.

Two definitions of the tunneling time delay are discussed in strong field ionization, which are related to the Eisenbud-Wigner time. The first is the asymptotic time delay (ATD), which is investigated in the attoclock experiments and the second is the time delay near the tunnel exit, exit time delay (ETD), which is only observable in a \textit{Gedankenexperiment} with a  virtual detector \cite{Feuerstein_2003,Wang_2013}. The signal of the virtual detector can be derived from the numerical solution of time-dependent Schr\"odinger equation (TDSE) calculating the  current density of the tunneled electron wave packet. The time delay at the tunneling exit calculated numerically with the virtual detector method~\cite{Yakaboylu_2013,Teeny_2016a,Teeny_2016b,Camus_2017} has been shown to be positive.

The ATD is extracted from the photoelectron momentum distribution (PMD) as a shift of PMD with respect to the expected distribution with the assumption of a vanishing time delay. The extraction of the ATD can be implemented using the method of the  classical backpropagation \cite{Ni_2016,Ni_2018b,Ni_2018a,Hofmann_2021}.  In the deep tunneling  regime  the ATD is vanishing \cite{Yakaboylu_2013, Han_2019}. However, near the over-the-barrier ionization (OTBI) regime it is nonnegligible and negative  \cite{Torlina_2015,Ni_2018b,Ni_2018a,Klaiber_2022}. This negative ATD is explained as arising due to interference of direct and the under-the-barrier recolliding trajectories \cite{Klaiber_2018}.  Note that the depletion of the bound state also induces a negative time delay. When investigating the time delay problem experimentally or via the numerical solution of the time-dependent Schr\"odinger equation, the depletion effect should be estimated separately and subtracted from the total time delay. In the strong field approximation (SFA) theory the depletion of the bound state is not included. This is an advantage offered by the SFA theory: the emergence of the time delay can be investigated avoiding any complications stemming from the depletion effect.

The notion of a third time delay related to tunneling ionization,  physically different from ATD and ETD, can be deduced from a recent experiment \cite{Yu_2022}. It is devoted to measuring the tunneling time delay in a setup where the attoclock is augmented by a second harmonic laser field of linear polarization. In this setup, the ionization yield is modulated with respect to variation in the time delay between the two components of the laser field. The experiment showed that the yield is maximal at vanishing time delay between the field components, i.e. when the total field of the two-color laser field has the largest maximum. From the latter it has been concluded that the tunneling time is vanishing. We analyzed the given experimental results to clarify the physical interpretation of the obtained vanishing tunneling time.
%The result of the experiment is that the peak of PMD is the largest at a such time delay between the two components of the laser fields when the total field of the two-color laser field has the largest maximum.
Our analysis within the SFA  model confirms the experimental finding that the PMD peak is the largest when the maximum of the total field is the largest.
%corresponds to the ionization wave packet emerging at the maximum of the total field.
At the same time, we confirm that the PMD shift due to the sub-barrier recollision, or the equivalent ATD, still is available. Therefore, we conclude that the vanishing time delay measured in the experiment of Ref.~\cite{Yu_2022} provides no information on ATD, although it is derived from the asymptotic PMD.

\begin{figure}[b]
\includegraphics[width=0.5\textwidth]{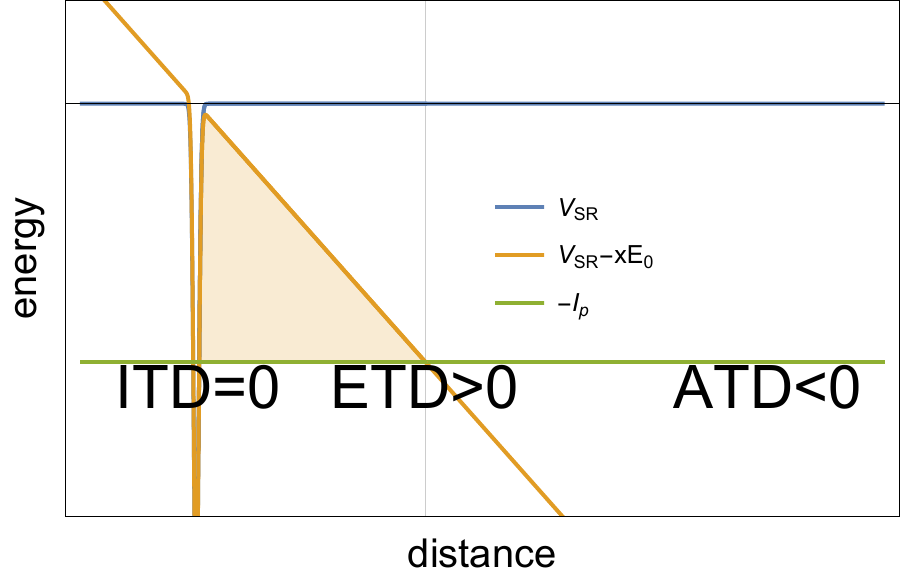}
 \caption{  Time delays related to the tunneling ionization: the initial time delay (ITD), describing the emergence of tunneling ionization wavepacket inside the  barrier (it is vanishing), the exit time delay (ETD), describing the peak of the tunneling wave packet near the tunnel exit (it is positive); asymptotic time delay (ATD) deduced from the asymptotic PMD (it is negative); $V_{SR}$ is the short-range atomic potential, $V_{SR}-xE_0$ is the tunneling ionization barrier through the laser field modified atomic potential, $-I_p$ is the bound state energy.}\label{scheme}
 \end{figure}

Can one introduce a time delay that describes the result of this experiment? To this end let us recall the quantum orbit picture of strong field ionization \cite{Becker_2002}, where ionizing complex trajectories start in the under-the-barrier region at a complex time $t_s$,   asymptotically becoming classical trajectories. One can call $t_s(\mathbf{p})$ as the time of initiation of the ionization wave packet around the asymptotic momentum $\mathbf{p}$. The delay of ${\rm Re} \{t_s(\mathbf{p}_m)\}$ corresponding to the peak of PMD $\mathbf{p}=\mathbf{p}_m$ with respect to the peak of the laser field, we may define as the tunneling initiation time delay (ITD). With this definition, the result of Ref.~~\cite{Yu_2022} can be interpreted as the measurement of  ITD to be vanishing, see the tunneling time delay scheme in Fig.~\ref{scheme}. Retrospectively, we can interpret the trajectory-free tunneling time calculated in Ref.~\cite{Eicke_2018}, using the saddle-point of the numerical solution of TDSE expressed via the Green function, as the calculation of ITD, and showing it to be vanishing, similar to the experimental result of Ref.~\cite{Yu_2022}. While ETD is determined by the peak of the electron wave packet near the tunnel exit, ATD -- by the shift of the peak of the asymptotic PMD, the ITD is determined by the emerging time  $t_s(\mathbf{p}_m)$ of the quantum orbit corresponding to the peak of PMD.

Recently, we have investigated the time delay in tunneling ionization using the first-order strong-field approximation (SFA) within the virtual detector approach \cite{Canario_2021}. The  calculation of the SFA wave function based on the direct ionization amplitude showed a positive time delay in the region of the  tunnel exit, without invoking recollisions. It has been confirmed that reflections of the electron wavepacket under the tunneling barrier are  responsible for this non-zero time delay around the tunnel exit.  While the ETD is not directly measurable in an experiment, it is amenable to measurement in a numerical experiment via a solution of the TDSE. In  \cite{Klaiber_2018} it has been clarified that the nonnegligible negative ATD  emerges due to the interference of the direct and sub-barrier recolliding paths, however, the relationship between asymptotic and exit time delays was not clear. In the present paper, we address the issue of the quantitative relationship between ATD and ETD.

In this paper we continue the investigation of the time delay in tunneling ionization within SFA. Our aim is to analyze the role of the under-the-barrier recollisions for the tunneling time delay at the tunnel exit. To this end we calculate the wavefunction of the tunneling electron with an accuracy up to the second-order SFA, which includes the recolliding quantum orbits. The Wigner trajectory is constructed corresponding to the peak of the ionized part of the wave function, and the time delay is extracted from the latter. While already the Wigner trajectory based on the first-order SFA wavefunction shows a positive ETD \cite{Canario_2021}, here we examine how it is perturbed by the sub-barrier recollisions.
Firstly, we employ the same simple model for tunnel-ionization as in the previous study \cite{Canario_2021}, considering a one-dimensional (1D) atom, with an electron bound by a short-range potential, which is ionized by a half-cycle laser pulse.
This simple model contains major features of the tunneling ionization and allows a fully analytical treatment. The 1D treatment is justified as the ionization occurs mainly in the direction of the electric field. The use of a half-cycle laser pulse excludes recollisions via the electron continuum dynamics, that are not related to the tunneling time delay. We consider the regime close to the OTBI regime. In this regime the ATD induced by sub-barrier recollisions is not vanishing, in contrast to the deep tunneling regime discussed in \cite{Canario_2021}. Secondly, we extend the 1D model into three dimensions keeping the short-range character of the binding potential,  and show that the qualitative features of the tunneling time delay are maintained in the 3D case.  Finally, we discuss the recent experiment \cite{Yu_2022} measuring the tunneling time delay in a two-color attoclock setup. Our  SFA  model confirms the experimental findings,  however, shows that the latter provides no information on the ATD, and a new definition of a time delay (ITD) is necessary to physically interpret the experimental results.

 The structure of the paper is as follows: in Sec.~\ref{section_ii} we introduce the theoretical approach based on the SFA formalism, in particular, the applied low-frequency approximation (LFA). The results for 1D and 3D cases are discussed in Sec.~\ref{section_iii}, the two-color experiment is analyzed in Sec.~\ref{2-color}, and our conclusion is given in Sec.~\ref{sec:conclusion}. Atomic units (a.u.) are used throughout the paper.

\section{Theory}\label{section_ii}

\subsection{Statement of the problem}

We consider strong-field ionization of an atom in a unipolar laser field, aiming at the investigation of the tunneling time delay. A short-range potential, $V(\br)$, is chosen to model an atomic potential. This allows the discussion to focus uniquely on the presence of time delay effects, unlike  angular streaking experiments for which long-range Coulomb effects must be accounted. The ionization process is described by the Schr\"odinger equation
\be
\label{schroeq}
i\frac{\partial}{\partial t}\Psi(\br,t)= (H_0 +H_i)\Psi(\br,t),
\ee
with the atomic Hamiltonian
\be
H_0 = -\frac {\boldsymbol{\nabla}^2}{2}+ V(\br),
\ee
and the interaction Hamiltonian of the electron with the time-dependent laser field $\mathbf{E}(t)$ in the length gauge
\be
H_i =   \br \cdot \bE(t)\;.
\ee
A unipolar laser pulse is employed to avoid multi-half-cycle interference effects, which could hinder the observation of the time-delay signature. The pulse is linearly polarized and has a Gaussian form:
\be
\bE(t)=-E_0 \exp\left[-(\omega\, t)^2\right] \hat{\mathbf{x}},
\ee
with the field strength $E_0$, and the angular frequency $\omega$. We are interested in the tunneling ionization regime, when the dimensionless Keldysh parameter \cite{Keldysh_1965} $\gamma=\omega \kappa/E_0\ll 1$
%\be \gamma=\sqrt{\frac{I_p}{2 U_p}}\ll 1, \ee
is much smaller than unity. Here, $-I_p=-\kappa^2/2$ is the atomic binding energy,
%and $U_p$  the ponderomotive potential which in a linearly polarized laser field is $U_p=E_0^2/(4\,\omega^2)$.
Throughout this paper we choose $\gamma=0.3$ and are, therefore, always in the quasistatic ionization regime,  while varying the laser field strength and, accordingly, the frequency.

\subsection{The Strong Field Approximation}

A formal solution to Eq. (\ref{schroeq}), is given by the time evolution operator $U(t,t_i)$ which unitarily evolves the wavefunction according to the full Hamiltonian $H=H_0+H_i$, from the initial state $|\psi(t_i\rangle$  at time $t_i$ into the state $|\psi(t)\rangle=U(t,t_i)|\psi(t_i)\rangle$ at time~$t$. In the interaction picture
the unitary time evolution operator  obeys  the Dyson equation
 \be
 \label{timeevol}
 U(t,t_i) = U_0(t,t_i) - i \int^t_{t_i} \: \diff t_1 \: U(t,t_1) H_i(t_1) U_0(t_1,t_i),
 \ee
 where $U_0$ is the evolution operator corresponding to the Hamiltonian $H_0$.
 We work within the well-known SFA \cite{Keldysh_1965,Faisal_1973,Reiss_1980}, which assumes that after ionization the electron dynamics are dominated by the laser field, treating any further interactions with the atomic core perturbatively. That is, the full time evolution operator is iterated with respect to the atomic potential $V $ \cite{Becker_2002}:
\begin{align}
\label{usfa}
U (t,t_1) = U_f(t,t_1)  - i \,\int_{t_1}^t \; \diff t_2\,U (t,t_2) \, V\, U_f(t_2,t_1) .
\end{align}
Here $U_f(t,t_1)$ is the time evolution operator for the electron dynamics purely in the laser field:
\be
U_f(t,t_1)= \intinfty \diff \bp \; \ket{\Psi_{\bp}(t)} \bra{\Psi_\bp(t_1)},
\ee
expressed via Volkov states of the electron in the laser field \cite{Volkov_1935},
 %\be
$\ket{\Psi_{\bp}(t)}=\ket{\bP(t)}  \exp[-i\,S_\bp(t)]$,
 %\ee
with the kinetic momentum $\bP(t)=\bp+ \bA(t)$, the phase $S_\bp(t)=\int^t d\tau \, \bP(\tau)^2/2$, and the laser vector potential $\bA(t)=\int^{\infty}_t \bE(\tau) \,\diff\tau$; the state $\braket{\br \,| \,\bp}=\exp\left[i \, \bp \cdot \br \right]/(2\pi)^{\frac{d}{2}}$ is the $d$-dimensional plane wave state.

Thus,  the electron wavefunction in the SFA is obtained as a truncated series with respect to the atomic potential $V$:
%Thus, formally, the electron wavefunction in the SFA is obtained as a truncated series in the evolution operator with respect to the atomic potential $V$, $U(t,t_i) = \sum_{n=0} U^{(n)}(t,t_i)$, and the full electron wavefunction takes the form of a series
\begin{equation}
\label{Psi}
  \ket{\Psi(t)}=\ket{\Psi_0(t)}+\ket{\Psi_i (t)} + \ket{\Psi_r(t)},
  %+ \mathcal{O}(V^2),
\end{equation}
 which describes the ionization of the ground state $|\Psi_0(t)\rangle$ via the direct path,
\be
\label{1sfa}
\ket{\Psi_i(t)}=-i \int^{t}_{t_i} \diff t_1\, \int \diff \bp  \ket{\Psi_\bp(t)}\braket{\Psi_\bp(t_1) \,| \,H_{i}(t_1) \,| \,\psi_0(t_1)},
\ee
 and via the recollison path, involving one further interaction with the atomic core, given by
\begin{eqnarray}
\label{2sfa}
\ket{\Psi_r(t)}&=&- \int^{t}_{t_i} \diff t_1  \int^{t}_{t_1} \diff t_2 \int \diff \bp \;  \int \diff \bk  \; \ket{\Psi_{\bp}(t)}\\[12pt]
&\times &\braket{\Psi_{\bp}(t_2) \,| \,V \,| \,\Psi_{\bk}(t_2)} \; \braket{\Psi_{\bk}(t_1) \, | \, H_i(t_1) \, | \, \psi_0(t_1)},\nonumber
\end{eqnarray}
with each further interaction with atomic potential corresponding to a higher order term in $V$.

The canonical interpretation of the above term, Eq.~(\ref{2sfa}), is that it corresponds to the event of an ionized electron being driven by the oscillating laser field back towards the core and scattering from it. However, in the case where the electron is not driven back towards its parent ion, in particular, in the case of the applied unipolar laser field, this term takes on a different meaning. In the quantum orbit picture, this additional perturbative interaction with the atomic core is described as a recollision in the complex (imaginary) time during the sub-barrier dynamics \cite{Milosevic_2014, Milosevic_2022}.

Such processes were dubbed \textit{``under-the-barrier recollisions"} in Ref.~\cite{Klaiber_2018}, where it was shown that, for ionization near the threshold for over-the-tunnelling-barrier ionization (OTBI), the interference between the direct ionization terms and under-the-barrier-recollision terms (i.e. between the first order term in the SFA, Eq.~(\ref{1sfa}) and higher orders in the SFA, e.g. Eq~(\ref{2sfa})) produced a measurable shift in the resulting asymptotic photoelectron momentum distribution (PMD).

The aim of the current work is  to investigate in which extent the time delay around the classical tunnel exit is altered due to the under-the-barrier recollisions, i.e., due to the higher order terms in the SFA. To this end the time dependent wave function $ \Psi(\mathbf{r},t) $ is calculated and the time delay around the classical tunnel exit is deduced via the maximum in time, $t$, of the spatial probability $|\Psi(\mathbf{r},t)|^2$ for a given coordinate, $\br$.

For comparison we also consider the asymptotic PMD  $|M(\bp)|^2$ measurable at a detector. This is calculated by the projection of the time-dependent wave function in the momentum representation
\begin{align}
m(\bp,t)=\braket{\Psi_\bp(t)\, |\,\Psi(t)},
\end{align}
on the free electron Volkov state $\ket{\Psi_\bp}$, in the limit of asymptotic times:
\begin{align}
\label{m}
M(\bp)= \lim_{t \rightarrow \infty} m(\bp,t).
\end{align}
In terms of the SFA  wave function of Eq.~(\ref{Psi}), the amplitude, $m(\bp,t)=m_1(\bp,t)+m_2(\bp,t)$ , with $m_1(\bp,t)=\braket{\Psi_\bp(t)\, |\,\Psi_i(t)}$, and $m_2(\bp,t)=\braket{\Psi_\bp(t)\, |\,\Psi_r(t)}$, which generates the asymptotic PMD  as a similar perturbation series
%\begin{align}
$M(\bp)=M_1(\bp)+M_2(\bp)$.
%+... \, ,\end{align}where the ground state contribution vanishes due to the orthogonality of the bound and free electron states.
The wave function via $m_1(\bp,t)$ we will call 1SFA wave function, while that via $m_1(\bp,t)+m_2(\bp,t)$, 2SFA one.

\subsection{The Low Frequency Approximation}

In the second-order SFA, recollisions are treated in the first-order Born approximation, which is inaccurate for intermediate energy electrons $|\bp|\sim\kappa$. For the under-the-barrier recollisons, given by Eq.~(\ref{2sfa}), exactly this condition applies and so the recollision treatment must be improved.

In order to do this, we make use of LFA \cite{Cerkic_2009,Milosevic_2014a}. In the LFA, the recollision matrix element in the Born approximation is replaced by the exact $T(\mathbf{p})$-matrix for the field free scattering off the zero-range potential:
\begin{align}
\braket{\Psi_\bp (t_2)\,|\, V \,| \,\Psi_{\bk}(t_2)} \rightarrow\ \braket{\Psi_\bp(t_2) \, | \, T(\bp+\bA(t_2))\, | \, \Psi_{\bk}(t_2)}.
\end{align}
In the LFA, the amplitude including recollisions with the core is then approximated by the integral:
\begin{align}
\label{2sfaR}
\ket{\Psi_r(t)}&=- \int^{t}_{t_i} \diff t_1 \int^{t}_{t_1} \diff t_2\int \diff \bp \int \diff \bk  \; \ket{\Psi_{\bp}(t)}\\[12pt]
&\times \braket{\Psi_{\bp}(t_2) \,|\, T(\bp+\bA(t_2))\,|\, \Psi_{\bk}(t_2)} \; \braket{\Psi_{\bk}(t_1) \,| \,H_i(t_1) \,| \psi_0(t_1)}.\nonumber
\end{align}

We simplify more LFA by correcting the second-order SFA by the so-called LFA-factor. This factor provides an analytical estimate of the effect of the LFA in the quasistatic limit. By  considering a constant electric field $\bE=-E_0\hat{\mathbf{x}}$, we derive the LFA-factor via the ratio of the second-order momentum amplitude in the LFA  to that in the second-order SFA:
\be
{\cal T}_{LFA}=\frac{m_{2}\,\Big\rvert_{\,LFA}}{m_2\Big\rvert_{\,SFA}}.
\ee
This is a relatively straightforward calculation, details of which can be found in Appendix~\ref{app_lfa}. In one dimension this ratio is calculated to be
\be
{\cal T}^{(1D)}_{LFA}=1-\sqrt{\frac{\pi}{2}\frac{\kappa^3}{E_0}}.
\ee

In three dimensions, the calculation contains additional integrations on the transversal coordinate and momentum. Under the assumption that the final momentum lays in the polarization axis $p_y=p_z=0$, we obtain a similar scaling
\be
{\cal T}^{(3D)}_{LFA}=\sqrt{\frac{\pi}{2}\frac{\kappa^3}{E_0}}.
\ee
The LFA-factor is then inserted as a prefactor into the time dependent second-order SFA amplitude
and the latter is calculated numerically in the quasi-static regime of $\gamma=0.3$.

\subsection{The electron wavefunction in momentum space}

The electron wavefunction in momentum space $m(\bp,t)= m_1(\bp,t)+m_2(\bp,t)$ reads in LFA:
\begin{align}
m_1(\bp,t)=& -i\int^{t}_{t_i} dt_1 \braket{ \Psi_\mathbf{p}(t_1)|H_i(t_1)|\psi_0(t_1)}\\
m_2(\bp,t)=& -{\cal T}_{LFA}\int^{t}_{t_i} dt_1 \int^{t}_{t_1}dt_2\int^{\infty}_{-\infty} d\mathbf{k}\, \braket{ \Psi_\mathbf{p}(t_2)|V(\mathbf{r})|\Psi_\mathbf{k}(t_2)}\\
\nonumber & \times \braket{ \Psi_\mathbf{k}(t_1)|H_i(t_1)|\psi_0(t_1)}
\end{align}

%\subsubsection{1D wavefunction}

In the 1D case the first-order amplitude can be expressed as
\begin{align}
\label{m1SFA}
m^{(1D)}_{1}(p,t)=&\int^{t}_{t_i} dt_1 \int^\infty_{-\infty} dx_1 \;\tilde{m}_1^{(1D)}(p, x_1,t_1),
\end{align}
 where for convenience we define the integrand of the matrix element
\be
\label{m1xt}
\tilde{m}_1^{(1D)}(p, x_1,t_1)=   -i \Psi^*_p(x_1,t_1)H_i(t_1)\psi_0(x_1,t_1).
\ee
The second-order amplitude can be simplified via analytical integration of the recollision coordinate $x_2$ in the first matrix element and the intermediate momentum $k$, which yields
\begin{align}
\label{mLFA}
m^{(1D)}_{2}(p,t)&=   \int^t_{t_i} dt_1 \int^{t}_{t_1}dt_2\intinfty dx_1\,\tilde{m}^{(1D)}_{2}(p,x_1,t_1,t_2)
\intertext{
with}
 \tilde{m}^{(1D)}_{2}(p,x_1,t_1,t_2&)=-i{\cal T}^{1D}_{LFA}  \;\sqrt{\frac{2\pi}{i(t_2-t_1)}} \nonumber \\
 &\times \Psi^*_p(0,t_2)(-\kappa)\Psi_{k_s}(0,t_2)\; \tilde{m}_1^{(1D)}(k_s, x_1,t_1). \label{m2xt}
\end{align}
where $k_s=(-x_1-\alpha(t_2)+\alpha(t_1))/(t_2-t_1)$ and $\alpha(t)=\int^t \;d\tau \;A(\tau)$.
Both amplitudes now are expressed via the integral
 \be
{\cal I}_i= \intinfty \, d x_1 \; x_1\; \exp\left[ a_i\,x_1 +b_i\, x_1^2 -\kappa \, \lvert \, x_1\, \rvert \right],\label{i11}
\ee
with corresponding coefficients $a_i$ and $b_i$ for $i=1,2$, shown in Table~\ref{table11}.
\begin{table}[h!]
\begin{center}
\begin{tabular}{c | c | c  }
\; &  $ {\cal I}_1$  &  ${\cal I}_2  $    \\[3pt]
\hline
 $a_i$\hspace{.25cm} \rule{0pt}{4ex} &$-i (p +A(t_1)) $&$-i(\tilde{k}_s+A(t_1))$  \\[10pt]
 $b_i$ \hspace{.25cm} &$ 0$& $\dfrac{i}{2(t_2-t_1)}$  \\[10pt]
\end{tabular}
\caption{Coefficients of the integral Eq. (\ref{i11}) for the first- and second-order SFA.
In the above $\tilde{k}_s=-(\alpha(t_2)-\alpha(t_1))/(t_2-t_1)$}
\label{table11}
\end{center}
\end{table}

The integral of Eq.~(\ref{i11}) has an analytic solution
\begin{eqnarray}
{\cal I}_i&=&\frac{\sqrt{\pi } e^{-\frac{\left( a_i + \kappa \right)^2}{4 b_i}}}{4 \left(-b_i\right)^{3/2}}\left\{e^{\frac{a_2 \kappa }{b_i}} \left(a_i-\kappa \right)
   \left[1+\text{erf}\left(\frac{a_i-\kappa }{2 \sqrt{-b_i}}\right)\right]\right.\nonumber\\
   &+&\left.\left(a_i+\kappa \right) \left(1-\text{erf}\left(\frac{a_i+\kappa }{2 \sqrt{-b_i}}\right) \right)\right\};\label{i2}
\end{eqnarray}
In the limit $b_i \rightarrow 0$, we have ${\cal I}_i=\frac{4a_i\kappa}{(a_i^2-\kappa^2)^2}$.

Finally, the amplitudes are calculated using the function ${\cal I}$ defined above:
\begin{eqnarray}
m^{(1D)}_{1}(p,t)&=&\int^{t}_{t_i} dt_1\,  \tilde{m}_1^{(1D)}(p,0,t_1) {\cal I}_1(t_1),\label{mm1}\\
m^{(1D)}_{2}(p,t)&=& \int^t_{t_i} dt_1 \int^{t}_{t_1}dt_2 \tilde{m}_2^{(1D)}(p,0,t_1,t_2)\, {\cal I}_2(t_1,t_2)\label{mm2}.
\end{eqnarray}

The derivation of the approximate SFA amplitudes in three dimensions is given in Appendix~\ref{app_3D_momentum}.

\subsection{The  electron wavefunction in coordinate space}
%\subsubsection{1D wavefunction}

The coordinate  wavefunction can be represented straightforwardly via a 1D-Fourier transformation using the SFA amplitudes $m_{1}$ and $m_2$  as follows:
\begin{eqnarray}
\Psi_i(x,t)&=& \int^{\infty}_{-\infty} dp\, m_{1}(p,t)\;\Psi_{p}(x,t)\label{wf1}\\
\Psi_r(x,t)&=&\int^{\infty}_{-\infty} dp\, m_{2}(p,t)\;\Psi_{p}(x,t)\label{wf2}
\end{eqnarray}
We underline that  the amplitudes $m_{1,2}(p,t)$ in the equations above are time-dependent. For the calculation of ATD we use $ M_{1,2}(p) =m_{1,2}(p,t)|_{t\rightarrow \infty}$, while for the exit delay $m_{1,2}(p,t)|_{t=t_e} $, at the exit time $t_e $ (close to zero), is employed.
The momentum integration is performed by SPA  yielding an extra factor:
\begin{eqnarray}
\Psi_i(x,t)&=&  \int^{t}_{t_i} dt_1 \int^\infty_{-\infty} dx_1\,\tilde{m}_1^{(1D)}(p_{s1}, x_1,t_1) \sqrt{\frac{2\pi}{i(t-t_1)}} \Psi_{p_{s1}}(x,t)\nonumber\\
\label{wf11}\\
\Psi_r(x,t)&=&\int^t_{t_i} dt_1 \int^{t}_{t_1}dt_2\intinfty dx_1\,\tilde{m}^{(1D)}_{2}(p_{s2},x_1,t_1,t_2)\nonumber\\
&\times& \sqrt{\frac{2\pi}{i(t-t_2)}} \Psi_{p_{s2}}(x,t),\label{wf22}
 \end{eqnarray}
where $p_{s1}=[x-x_1-\alpha(t)+\alpha(t_1)]/(t-t_1)$, and $p_{s2}=[x-\alpha(t)+\alpha(t_2)]/(t-t_2)$. The coordinate integration can be represented again by the functions ${\cal I}_i$, with the coefficients $a_i$ in the integral  of Eq. (\ref{i11}) are now given by those of Table~ \ref{table2}.
\begin{table}[h!]
\begin{center}
\begin{tabular}{c | c | c }
\; &  ${\cal I}_1$  &  ${\cal I}_2$\\[3pt]
\hline
 $a_1$\hspace{.25cm} \rule{0pt}{4ex} &$-i (\tilde{p}_{s1} +A(t_1)) $ & $-i(\tilde{k}_s+A(t_1))$   \\[10pt]
 $b_i$ \hspace{.25cm} &  $ \dfrac{i}{2(t-t_1)}$& $\dfrac{i}{2(t_2-t_1)}$\\[10pt]
\end{tabular}
\caption{Coefficients of the integral Eq. (\ref{i11}) for the first- and second-order SFA wavefunction in the coordinate representation.
In the above $\tilde{p}_{s1}=(x-\alpha(t)+\alpha(t_1))/(t-t_1)$, $\tilde{k}_s=-(\alpha(t_2)-\alpha(t_1))/(t_2-t_1)$.}
\label{table2}
\end{center}
\end{table}
\begin{eqnarray}
 \Psi_i(x,t)&=&\int^{t}_{t_i} dt_1\,  \tilde{m}_1^{(1D)}(\tilde{p}_{s1},0,t_1) {\cal I}_1(t_1)\sqrt{\frac{2\pi}{i(t-t_1)}} \Psi_{\tilde{p}_{s1}}(x,t),\nonumber\\
 \label{mmm1}\\
 \Psi_r(x,t)&=&\int^t_{t_i} dt_1 \int^{t}_{t_1}dt_2 \tilde{m}_2^{(1D)}(\tilde{p}_{s2},0,t_1,t_2)\, {\cal I}_2(t_1,t_2)\nonumber\\
 &&\times \sqrt{\frac{2\pi}{i(t-t_2)}} \Psi_{\tilde{p}_{s2}}(x,t),\label{mmm2}
\end{eqnarray}
with $\tilde{p}_{s1}=(x-\alpha(t)+\alpha(t_1))/(t-t_1)$, and  $\tilde{p}_{s2}=(x-\alpha(t)+\alpha(t_2))/(t-t_2)$.
Further, the probability distribution at the exit, $x=x_e$, is calculated via the wavefunction $|\Psi(x_e,t)|^2$.
Here, the average exit coordinate $x_e$  is obtained by averaging over the tunneling probability (Keldysh-exponent), similar to Ref.~\cite{Canario_2021}:
\begin{eqnarray}
\label{exit_coo}
x_e=\dfrac{\intinfty dt \; \frac{I_p}{|\bE(t)|}\exp[-\frac{2\kappa^3}{3|\bE(t)|}]}{\intinfty dt \exp[-\frac{2\kappa^3}{3|\bE(t)|}]}.
\end{eqnarray}

The 3D wavefunction in coordinate space is given in Appendix~\ref{app_3D_coordinate}.

\begin{figure}[b]
\includegraphics[scale=1]{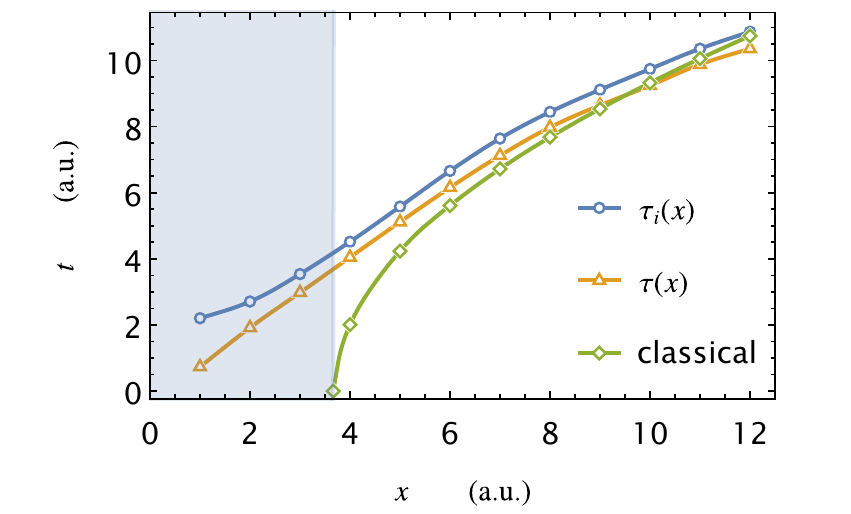}
 \caption{Wigner trajectories in the 1D case: (orange triangles) the trajectory $\tau (x)$ calculated via the  2SFA wave function including the direct and sub-barrier recolliding paths; (blue circles) the trajectory $\tau_i (x)$ calculated via the 1SFA wave function including only the direct ionization path; (green diamonds) the classical trajectory starting at the tunnel exit with a vanishing velocity. The shaded area indicates regions under the barrier, i.e. smaller than the tunnel exit coordinate $x_e$ given by Eq.~(\ref{exit_coo}). The field strength $E_0=0.15$ a.u. is below the OTBI threshold.}\label{trajectory}
 \end{figure}

 \section{Discussion}\label{section_iii}

 \subsection{The Wigner Trajectory}

\begin{figure*}[t]
\subfloat[\label{subfig:2a}]{
\includegraphics[scale=1]{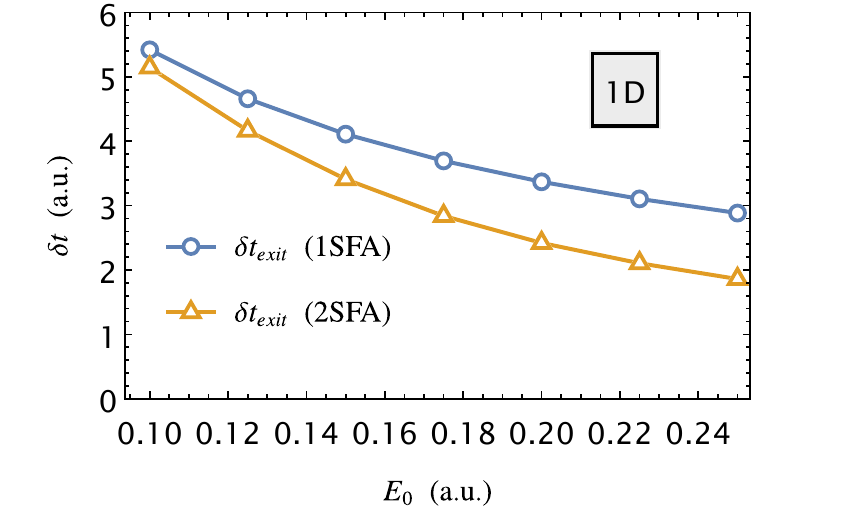}
}
\subfloat[\label{subfig:2b}]{
\includegraphics[scale=1]{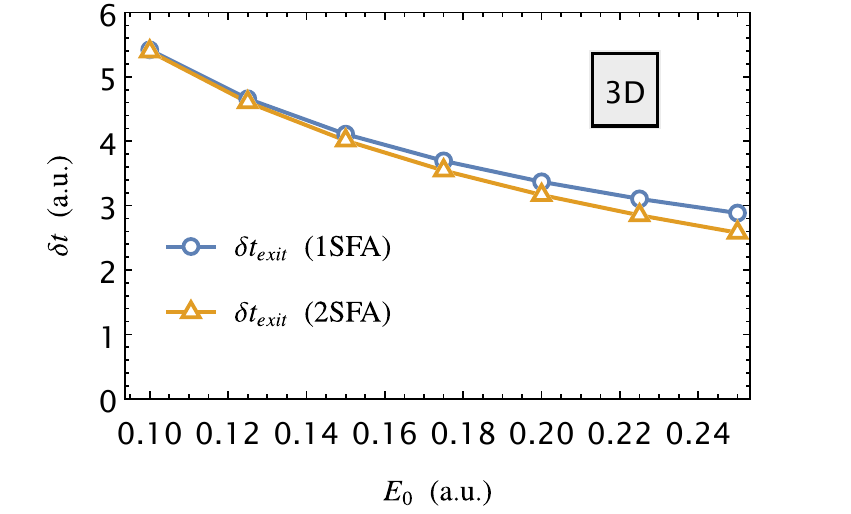}
}
 \caption{Time delay at the tunnel exit vs the laser field: (a) in the 1D case, (b) in the 3D case, using the first order SFA (blue circles), as well as the second order SFA (orange triangles). Both time delays $\delta t_{exit} = \textnormal{max}_t[\mathcal{P} (x_e,t)]- \textnormal{max}_t[E(t)]$, are calculated as the peak of the temporal probability distribution at the tunnel exit $x_e$, as laid out in the text. In the 2SFA (orange triangles), the probability ${\cal P}(x,t)= |\Psi_i(x,t)+\Psi_r(x,t)|^2$ includes the effects of recollisions, whereas as in the 1SFA the probability distribution ${\cal P}_i(x,t) =|\Psi_i(x,t)|^2$ accounts only for direct ionization. The difference of these two delays, $|\delta t_{exit}^{(r)}|= |\delta t_{exit}(2SFA)-\delta t_{exit}(1SFA)|$ increases with field strength, an effect directly attributable to the under-the-barrier recollision.}
 \label{ETD}
 \end{figure*}

 \begin{figure}[b]
\includegraphics[width=0.5\textwidth]{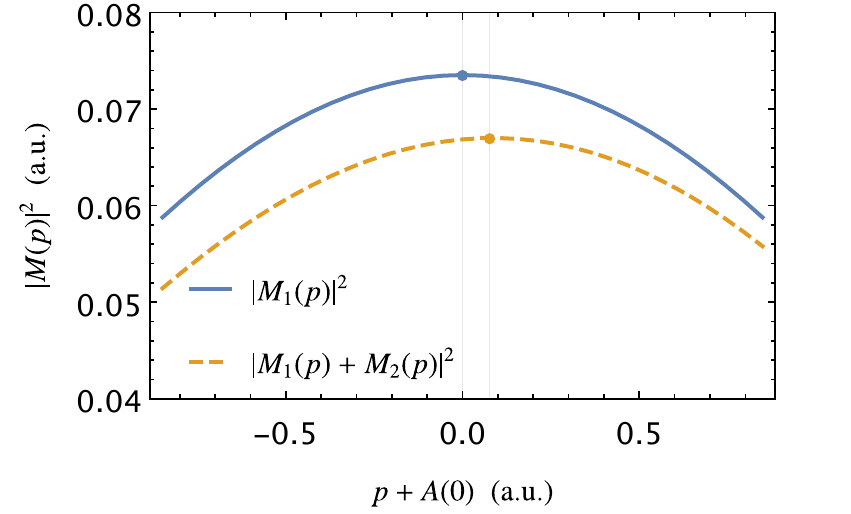}
 \caption{Asymptotic momentum distribution in 1D for the wavefunction using the SFA up to first (blue, solid) and second order (orange, dashed).  A positive momentum shift $\delta p \approx 0.08$ a.u. is observed in the peak of the distribution when one under-the-barrier recollison is considered, corresponding to a negative time delay of $-0.53$ a.u. The grid lines, and associated coloured dots, indicate the peaks of the distributions.}\label{PMD}
 \end{figure}

   \begin{figure*}
\subfloat[\label{subfig:5a}]{
\includegraphics[scale=1]{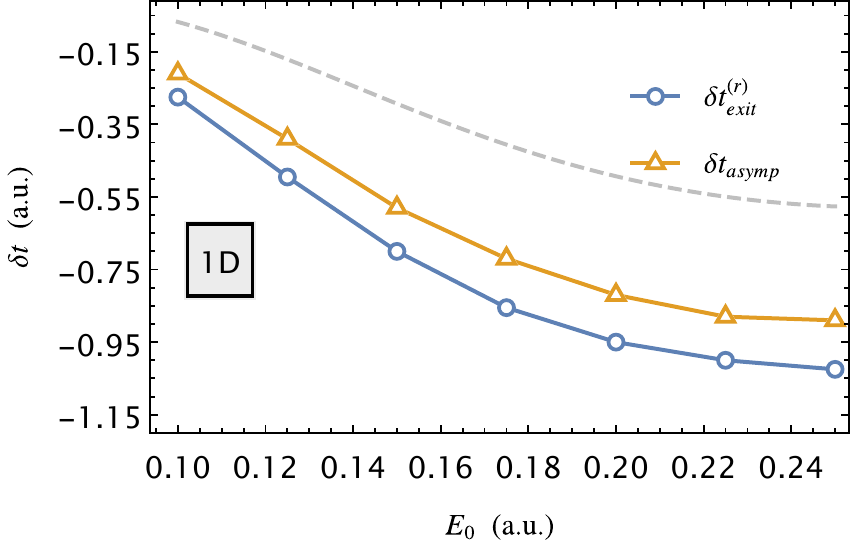}
}
\subfloat[\label{subfig:5b}]{
\includegraphics[scale=1]{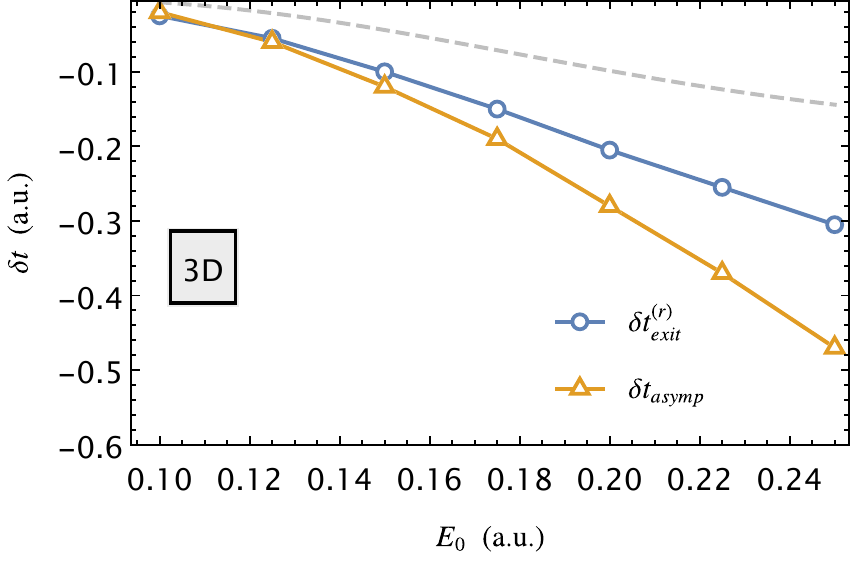}
}
 \caption{ Relationship between ETD and ATD in the (a) 1D and (b) 3D cases. Plotted are the variation of the peak of the tunnelling wave packet at the tunnel exit due to the recollision, $\delta t_{exit}^{(r)}= \delta t_{exit}(1SFA)-\delta t_{exit}(2SFA)$ (orange triangles), ATD $\delta t_{asymp}=\delta p_{asymp}/E_0$ (blue circles). The dashed line shows an estimate of ATD in a static field calculated via the Wigner derivative and given by Eq.~(\ref{qsWe}) and its 3D equivalent.}
   \label{ATD}
 \end{figure*}

The dynamics of the laser driven electron during strong-field ionization are described by the SFA wavefunction $\Psi_i(x,t)+\Psi_r(x,t)$. The Wigner trajectory $\tau_W(x)$ is derived using the  probability ${\cal P}(x,t)= |\Psi_i(x,t)+\Psi_r(x,t)|^2$  of the laser driven wavepacket
in the following way: for each fixed space point $x$, $\tau_W(x)$ corresponds to the peak of the  probability ${\cal P}(x,t) $.  For comparison, we additionally consider an analogous ``direct ionization'' Wigner trajectory, $\tau^{(i)}_{W}(x)$, calculated only using the maximum of the direct ionization probability ${\cal P}_i(x,t)=|\Psi_i(x,t)|^2$.  These Wigner trajectories for the field strength $E_0=0.15$ a.u. are shown in Fig.~\ref{trajectory}, beginning at the typical value $|x_s|\sim 1/\kappa$, which is the $x_1$-saddle point of the product of the 1D bound state $\sim\exp[-\kappa|x_1|]$ with the interaction Hamiltonian $H_i\sim E(t)x_1$. The field value is chosen not to exceed, but to be close to the threshold for OTBI, when the tunneling time delay is significant.

Both the direct ionization Wigner trajectory, $\tau^{(i)}_{W}(x)$, and the one via the full ionization amplitude including a recollision, $\tau_W(x)$, show a positive time delay at the tunnel exit compared to the peak of the laser field. However, the recollision under the barrier acts to reduce the time delay slightly. In Ref.~\cite{Canario_2021} we have shown that the positive time delay of the direct ionization path arises due to reflections inside the barrier, and it is positive as the reflections hinder the wavepacket crossing the barrier.  The positive ETD is reduced by the sub-barrier recollision, which can be intuitively explained by the additional positive probability current induced by the recollision. That is, accounting for the additional possibility of ionization through a recollision increases the probability current by accounting for an additional ionization channel. An increase in probability flux implies reduced hindrance of the tunnelling wavefunction which in turn implies a smaller time delay.

Far from the exit,  the direct trajectory  approaches the classical trajectory, i.e. the trajectory of a  classical electron appearing at $x_e$ at the peak of the laser field, with a vanishing momentum. Thus, the direct trajectory  shows vanishing ATD with respect to the ``simple man'' model (tunneling, followed by classical motion), while the trajectory containing an under-the-barrier recollision tends to a negative ATD with respect to the simple man model. The latter is in accordance with the previous result of Ref.~\cite{Klaiber_2018}.

\subsection{Time Delay Dependence on Field Strength}

The dependence of ETD, $\delta t_{exit}$, on the laser field strength, $E_0$, is shown in Fig.~\ref{ETD}. With larger fields, the time delay decreases, which was already established for direct ionization in Ref.~\cite{Canario_2021}.

However, the effect of recollisions on the time delay, i.e., the difference of the time delay between the direct and recolliding trajectories,
\be
\delta t_{exit}^{(r)}=\delta t_{exit}(2SFA)-\delta t_{exit}(1SFA),
\ee
in this case increases by absolute value, as shown in Fig.~\ref{ETD}.

Thus, on the one hand, the sub-barrier recollision  decreases  the positive ETD, i.e. has a negative contribution  to the ETD. On the other hand, it is known \cite{Klaiber_2018,Klaiber_2022} that  the sub-barrier recollision also induces a shift of the peak
\be
\delta p_{asym}=\textnormal{max}_p \{|M_1(p)+M_2(p)|^2\}-\textnormal{max}_p\{|M_1(p)|^2\}
 \ee
of the asymptotic PMD,  corresponding to a negative ATD
\be
\delta t_{asym}=-\delta p_{asym}/E_0.
\ee

We illustrate the latter in Fig.~\ref{PMD}, where the asymptotic PMD, $|M_1(\bp)+M_2(\bp)|^2$, via the SFA up to first  and  second  orders is shown for a field strength $E_0=0.15 \au$. A positive momentum shift $\delta p\approx 0.08$ a.u. (corresponding to the negative time delay $\delta t_{asym}\approx -0.53$ a.u.) is observed in the peak of asymptotic momentum, which is directly attributable to the under-the-barrier recollison.

\begin{table*}[t]
\begin{center}
\begin{tabular}{c| c | c | c| c}
\;& Simple man model & \quad1SFA \quad &  2SFA  & characteristic values \\
\hline
ITD &0 & 0& 0&0 \\[8pt]
ETD&0 &$E_0^{\; -2/3}$ &\quad $E_0^{\; -2/3}-\tau_0\exp(-\frac{2\kappa^3}{3E_0})$ \quad &$\lesssim 10$ a.u.\\[8pt]
ATD&0 &$0$& $-\tau_0\exp(-\frac{2\kappa^3}{3E_0})$&$\gtrsim -1$ a.u.\\[8pt]
Asymptotic momentum shift&0 &$0$& $E_0 \tau_0\exp(-\frac{2\kappa^3}{3E_0})$&$\lesssim 0.4$ a.u.\\[8pt]

\end{tabular}
\end{center}
\label{tabel2}
\caption{Scaling of tunnelling times and asymptotic momentum w.r.t field strength, $E_0$, for $E_0\ll \kappa^3$, for three different models of ionization in one dimension (the classical simple man model, direct ionization using the first order SFA, and ionization including one recollision using the second order SFA). These estimates provide upper or lower bounds for these measures which are indicated above as characteristic values. Here $\tau_0=\frac{\sqrt{2\pi}}{\kappa^2}\left(\frac{\kappa^3}{E_0}\right)^{3/2}$, as in Eq.~(\ref{qsWe2}). For the scaling via 1SFA see Ref.~\cite{Canario_2021}.}
\label{T3}
\end{table*}

We can give an estimate of the scaling of $\delta t_{asym}$  with respect to the field strength by calculating the Wigner time delay\cite{wigner_1955} of an electron in an adiabatic field \cite{Klaiber_2013,Yakaboylu_2013}. The Wigner delay corresponds to the energy derivative,
\be
\label{wigner}
 \delta t_{asym}=i\frac{\partial \ln(\Psi_\kappa)}{\partial I_p},
\ee
of the electron wavefunction in a static field, which in the SFA reads
\be
\Psi_\kappa \sim \exp\{-\kappa^3/(3E_0)\}+i \, {\cal T}^{(1D)}_{LFA} \,\exp\{-\kappa^3/E_0\}.
\ee
This equation has a simple intuitive explanation. The first term ($\exp\{-E_a/(3E_0)\}$) describes the direct tunneling amplitude and  its module square ($\exp\{-2E_a/(3E_0)\}$) is proportional to the tunneling exponent of Keldysh theory. The second term describes the recolliding path, which includes triple tunneling through the barrier: from the atom to the surface of the barrier, tunneling again toward the atom with a recollision, and the final tunneling leading to ionization.  Due to the triple tunneling, the tunneling exponential factor is repeated three times ($\exp\{-E_a/E_0\}$). The recollision is included via perturbation theory, therefore, the amplitude is proportional to the scattering amplitude by the core (${\mathcal T}_{LFA}^{(1D)}$). For an explicit derivation, we refer the reader to Eq.(2) of Ref.~\cite{Klaiber_2018}.
 Thus, a straight forward calculation (recalling the binding energy $-I_p=-\kappa^2/2$) yields
\begin{align}
\label{qsWe}
|\delta t_{asym|}&\sim  \dfrac{e^{-\frac{2\,\kappa^3}{3\,E_0}}}{\kappa^2} \left[-3\sqrt{\frac{\pi}{8}} \left(\frac{\kappa^3}{E_0} \right)^{\frac12} -2 \left(\frac{\kappa^3}{E_0} \right) +\sqrt{2\pi}  \left(\frac{\kappa^3}{E_0} \right)^{\frac32}\right]
\end{align}
which for $E_0\ll \kappa^3$ is dominated by the last term
\begin{align}
|\delta t_{asym}|&\sim \sqrt{2\pi}\frac{e^{-\frac{2\,\kappa^3}{3\,E_0}}}{\kappa^{2}}\left(\dfrac{\kappa^{3}}{E_0}\right)^{3/2}.
\label{qsWe2}
\end{align}

A similar derivation yields a three dimensional estimate $\delta t_{asym}^{(3D)} \sim (E_0/\kappa^3) \; \delta t_{asym}$, when the 3D SFA wavefunction is employed, $\psi^{(3D)}_\kappa\sim  \exp\{-\kappa^3/(3E_0)\}+i \, c_2(0, t_{s1},t_{s2}) \;{\cal T}^{(3D)}_{LFA} \,\exp\{-\kappa^3/E_0\}$, where $c_2(0, t_{s1},t_{s2})$ is given by Eq.~(\ref{c2xtt}), $t_{s1}= 3\,  t_{s2} = 3 i\kappa/E_0$ are the saddle points of time integration in a constant field, for details see Ref.~\cite{Klaiber_2018}.

These estimates are in good qualitative accordance with the time dependent SFA results shown in Fig.~\ref{ATD}.

\subsection{Relationship Between Asymptotic and Exit Delays}

It is interesting to see whether there is a relationship between these two time delays, $\delta t_{asym}$ and $\delta t_{exit}$. This question is particularly  relevant to attosecond streaking techniques which attempt to extract information on the tunnelling process from measurements at distances much greater than the atomic scale.
This question is analyzed in Fig.~\ref{ATD}, comparing these two time delays.

The qualitative features of the ETD are similar in both the 1D and 3D cases. The ETD is positive and decreases with larger fields. The sub-barrier recollision reduces the ETD [Fig.~\ref{ETD}], and the sub-barrier recollision effect increases with the field [Fig.~\ref{ATD}]. The value of the time delay in the 3D case is smaller than in 1D, because of the decreasing contribution of the recolliding wave packet which spreads in three dimensions  for the case of a short-range atomic potential.

\section{Tunneling time delay in a two-color laser field}\label{2-color}

Recently, a new scheme for the determination of the tunneling time has been proposed and experimentally implemented in Ref.~\cite{Yu_2022}. The new scheme uses an elliptically polarized infrared (IR) streaking laser field along with an additional perturbative second-harmonic field, linearly polarized along the major axis of the elliptical polarization. Due to the second harmonic field, the total laser field is slightly modified, which leads to slight modification of PMD.

\begin{figure*}
\subfloat[\label{subfig:6a}]{
\includegraphics[scale=0.8]{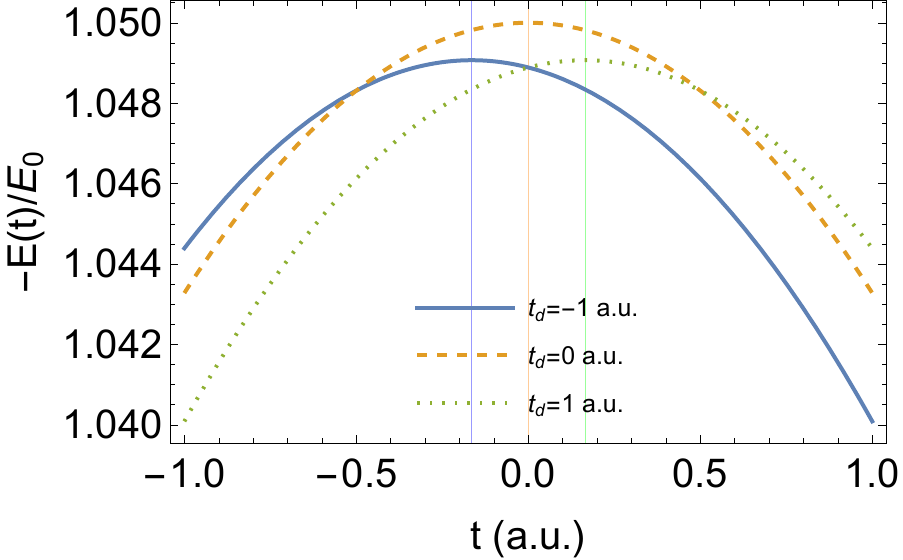}
}
\subfloat[\label{subfig:6b}]{
\includegraphics[scale=0.8]{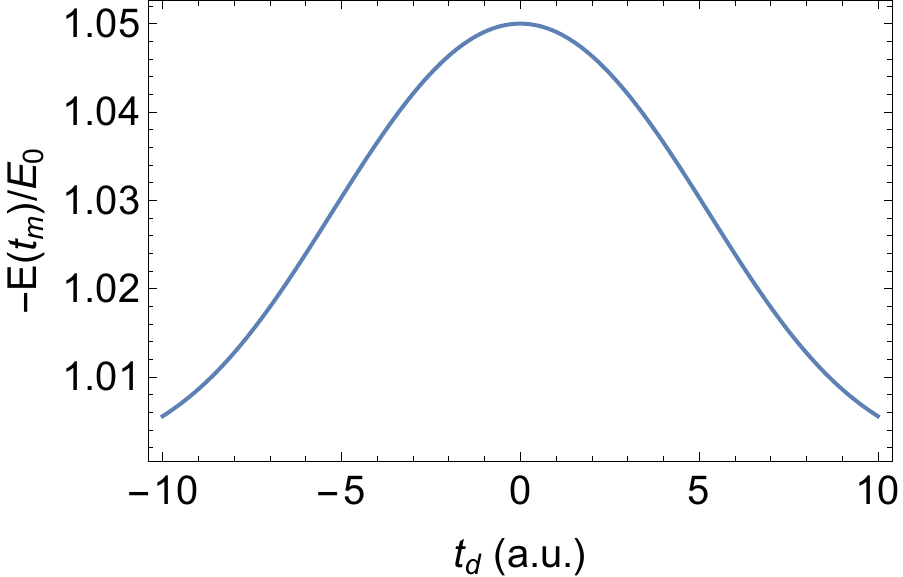}
}
 \caption{  The two-color field of Eq.~(\ref{color_field}): 
  (a) The total two-color field for $t_d=-1$ a.u. (blue solid), $t_d=0$ (orange dashed), and  $t_d= 1$ a.u. (green dotted); (b) The peak value of the two-color field $E(t_m)$  vs the time delay $t_d$.}
   \label{2-color-field}
 \end{figure*}

  \begin{figure*}
\subfloat[\label{subfig:7a}]{
\includegraphics[scale=1]{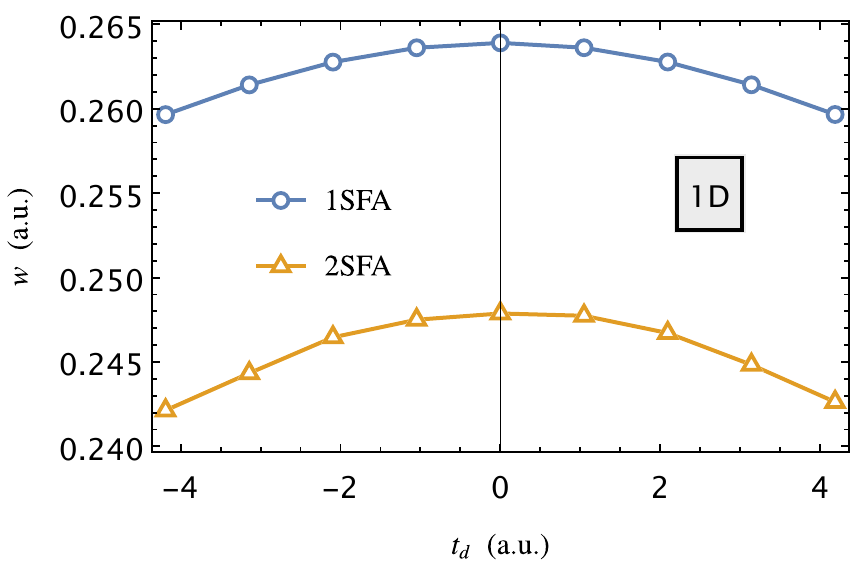}
}
\subfloat[\label{subfig:7b}]{
\includegraphics[scale=1]{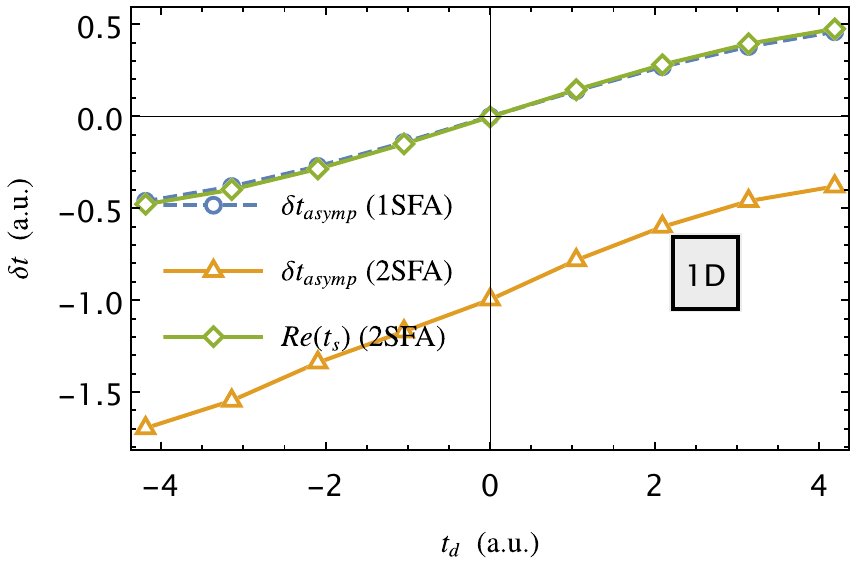}
}
 \caption{Tunneling ionization in a two-color laser field: (a) The probability $w$ for the peak of PMD vs the time delay $t_d$ between the color fields; (b) $\delta t_{asymp} = -(p_m+A(0))/E_0$ vs $t_d$, with the $p_m$ corresponding to the PMD peak. The ATD corresponds to the shift of  $\delta t_{asymp}$ between 2SFA and 1SFA, which is  due to the interference of the direct and the sub-barrier recolliding paths; (dashed line with blue cycles) via  1st-order SFA, (orange triangles) via 2nd order SFA, (green triangles)  ${\rm Re} \{t_s\} $ via 2SFA. The ITD corresponds to the difference between ${\rm Re}\{ t_s\}$ and 1SFA and it is vanishing for any $t_d$.  }
   \label{ITD}
 \end{figure*}

Due to the total field modification, the ionization yield at a given attoclock angle oscillates with respect to the phase difference (time delays) between the color fields. The experiment shows that the yield is the largest for the attoclock angle, which corresponds to the vanishing phase difference between the two color fields $\Delta\phi=\omega t_d=0$, with the time delay between the laser pulses. In the $\Delta\phi=0$ case, the peak field of the second-harmonic wave is added to that of the elliptical polarization, creating the largest total field inducing the ionization. Therefore, according to the experiment, the tunneling of the electron which produces the largest yield (largest  PMD peak), happens at the maximum of the field. From the latter, a conclusion is drawn that the tunneling time delay is vanishing. While the results of this accurate experimental measurement raise no doubt, the conclusion drawn from the results needs clarification. One needs to understand the physical meaning of the measured time delay.

%The modification of the total field can be most simply seen in the case of the  vanishing phase difference between the two colors $\Delta\phi=0$. In this case, during the laser period, the perturbative second harmonic field is aligned along the major axis of the large IR laser field ellipse, once increasing, and second time decreasing the total field.

We have tested the conclusion of these experimental results with our simple analytical 1D SFA model developed in previous sections, which uses a short-range potential for the atomic potential and two half-cycle laser pulses.
We mimic the two-color driving laser pulse with a field strength
\begin{eqnarray}
E(t) = -E_0\left\{\exp[(\omega t)^2]+\xi \exp[-(2(\omega(t - t_d)))^2]\right\},
\label{color_field}
\end{eqnarray}
which has the same property as the total field in the experimental setup: the total field is the largest when the time delay $t_d$ between the color field is vanishing, see Fig.~\ref{2-color-field}.

We calculate the PMD using Eqs.~(\ref{mm1})-(\ref{mm2}) (for the asymptotic time $t\rightarrow \infty$) as a function of the time delay $t_d$. The chosen parameters are $E_0 = 0.25$ a.u., $\omega = 0.075$ a.u. and $\xi = 0.05$. The probability  of the PMD peak varies depending on $t_d$ as is shown in Fig.~\ref{ITD}(a).  We observe the same result as in the experiment of Ref.~\cite{Yu_2022} that the probability of the PMD peak is the largest at vanishing time delay between the pulses, $t_d=0$, i.e.  when the field inducing ionization is the largest. Thus, the correspondence of the largest PMD peak to the largest field is confirmed, i.e., the largest PMD peak is initiated at the peak of the field.

We find from PMD the momentum $p_m$ corresponding to the PMD peak at any given $t_d$, and translate it to the ionization time  via the time delay $\delta t_{asymp} = -(p_m+A(0))/E_0$, taking into account that at $t=0$ the asymptotic momentum is $-A(0)$, with the vector potential $A(t)$. Note that $\delta t_{asymp}$ for 1SFA coincides with the simple man prediction for the peak momentum $p_m=-A(t_m)$, where $t_m$ corresponds to the peak of the field $E(t)$.

We calculate also the saddle-point time $t_s$ of the SFA amplitude, which corresponds to the time of the initiation of the quantum orbit of tunneling ionization. Both, 1SFA and 2SFA give the same value for ${\rm Re}\{ t_s\}$. Both, $\delta t_{asymp}$ and ${\rm Re}\{ t_s\}$ are shown in Fig.~\ref{ITD}(b), which indicates that  ${\rm Re}\{ t_s\}$ coincides with the 1SFA result, which is similar to the simple man model, i.e.  the ionization is initiated at the time of the maximum of the laser field. Thus, ITD corresponds to the difference between ${\rm Re}\{ t_s\}$ and 1SFA and it is vanishing for any $t_d$. The same conclusion that ${\rm Re}\{ t_s\}$ corresponding to the PMD peak is vanishing, has been derived in Ref.~\cite{Eicke_2018} via analysis of the numerical amplitude. The numerical solution of TDSE has been expressed in \cite{Eicke_2018} via the time integral containing the numerical Green function, and the saddle-point of this time integration has been obtained, identifying it with the tunneling time delay. We argue that in this way, in fact, the vanishing ITD has been calculated.

From Fig.~\ref{ITD}(b) we can deduce also the value of ATD, which corresponds to the shift of  $\delta t_{asymp}$ between 2SFA (full SFA amplitude, including the 1st and 2nd order SFA amplitudes) and 1SFA (1st order SFA). This shift is  due to the interference of the direct and the sub-barrier recolliding paths. The ATD (shift between 2SFA and 1SFA) is about 1 a.u. at any $t_d$. The ATD at $t_d=0$ (when the ionization yield is maximal) is $\delta t_{asymp}=-1$ a.u., while ITD via $ t_s$ is vanishing. Thus, in this two-color setup, the largest PMD peak of the ionization wave packet  originates at the peak of the laser field, $t_d=0$. The latter can be interpreted as  the ITD vanishing.

Further, there are two paths of ionization: the direct path, described by the 1st-order SFA amplitude, and the sub-barrier recolliding one, described by the 2nd-order SFA amplitude. While these paths originate at the peak of the laser field, their interference is observed in the PMD as a shift of the momentum distribution (with respect to the case only with the direct SFA path), which is equivalent to the nonzero ATD. Consequently, the emergence of the ionization wave packet at the peak of the laser field, i.e., the vanishing of ITD, does not preclude the nonvanishing ATD due to the sub-barrier dynamics.

 \section{Summary and conclusion}\label{sec:conclusion}

We have considered the tunneling time delay of an electron in strong-field ionization in a unipolar time-dependent laser field, accounting for under-the-tunneling-barrier processes. The electron wave function within a simplified model of ionization, with a short-range atomic potential, has been calculated analytically using SFA.
We considered the direct ionization (via the first-order SFA), and the full ionization amplitudes including the direct tunneling path and the path with the sub-barrier recollision (via the second-order SFA).
Employing the wavefunction in its spatial representation, we derived the Wigner trajectory near the tunnel exit. The Wigner trajectory shows a positive time delay near the tunnel exit both  with and without under the barrier processes. However, we find that when one accounts for sub-barrier recollisions the ETD is decreased slightly, see the summary in Table~\ref{T3}.

As is known from Ref.~\cite{Klaiber_2018}, the interference  of the direct and sub-barrier recolliding paths induces an asymptotic momentum distribution shift, which is equivalent to a negative time delay with respect to the simple man model. We found a relationship between the change of ETD due to sub-barrier recollisions and the ATD. Furthermore, we proved that these time delays are equal in the tunneling regime, as expected because of the same origin related to the effect of the sub-barrier recollision. The field dependence of these time delays is also obtained.

We provided also the 3D generalization of our results. The features of the tunneling time delay were shown to be similar to those in one dimension.

Finally, we tested with our model the conclusion drawn from the experiment on the accurate measurement of the ionization PMD in a two-color laser field. The result of the experiment shows that the ionization wave packet corresponding to the largest PMD peak emerges at the peak of the laser field. We introduced the notion of ITD and relate it to the two-color experiment, as well as to the theoretical calculation of the trajectory-free tunneling time of Ref.~\cite{Eicke_2018}. While the two-color experiment shows vanishing ITD,  this does not preclude, according to our theoretical analysis, the nonvanishing ATD due to the sub-barrier dynamics.

\section*{Acknowledgment} The discussions with Christoph Keitel are gratefully acknowledged. %We acknowledge the help of Daniel Bakucz Can\'{a}rio in numerical calculations and plotting the figures.

%\pagebreak

 \appendix

\section{The LFA factor}
\label{app_lfa}

\subsection{1D LFA Factor}
 We derive the LFA factor in the 1D case. To this end we  compare the second-order SFA amplitude with that of LFA in the quasistatic limit, namely, in a static field. The second-order SFA amplitude in a constant field $E(t)=-E_0$ has the following structure  in the 1D case after performing the coordinate integrations analytically:
\begin{eqnarray}
\label{1Dm2}
m_{2SFA}^{(1D)}&=&\int_{-\infty}^{\infty} dt_1 \int_{t_1}^{\infty} dt_2 \int dk\; {\cal P}_{2SFA} \\&& \exp\Big[-i\frac{p^2}{2}(t-t_2)-ip(\alpha(t)-\alpha(t_2))\nonumber\\
&&-i\frac{k^2}{2}\left(t_2-t_1\right)-ik\left(\alpha(t_2)-\alpha(t_1)\right)+i\beta(t_1)+i\frac{\kappa^2}{2}t_1\Big]\nonumber,
\end{eqnarray}
where $\alpha(t)=E_0t^2/2$ and $\beta(t)=E_0^2t^3/6$. In the above, ${\cal P}_{2SFA}=-\kappa^{5/2}/(\sqrt{2\pi})^{3}$ is the pre-exponential factor of the second-order  SFA, which we is weakly dependent on the integration variables. Without loss of generality we can set $p=0$ and simplify. Then, using SPA for the $k$- and $t_1$-integrations yields
\begin{align}
\label{m2sfa}
m_{2SFA}^{(1D)}= \int_{t_{1,s}}^{\infty} dt_2 \; \frac{2\pi}{\kappa}{\cal P}_{2SFA}  \exp\left[-\frac{\kappa^3}{E_0}-\frac{E_0\kappa}{2} \left(t_2 - i \frac{\kappa}{E_0}\right)^2\right],
\end{align}
where the exponent is already expanded quadratically  around the saddle point in $t_2$.  The integration contour in Eq.~(\ref{m2sfa}) consists of two parts: 1) from $t_{1,s}$ to $i\kappa/E_0$, and 2) from $i\kappa/E_0$ to $\infty$. The integration along the first part of the contour gives the direct ionization amplitude $m_1$, see Appendix \ref{A1},  cf. \cite{Lohr_1997}, which is dropped because in this section our aim is to derive the LFA factor for the recollision amplitude. Then, the integral along the second part of the contour yields
\begin{eqnarray}
m_{2SFA}^{(1D)}={\cal P}_{2SFA}\frac{(2\pi)^{3/2}}{2\kappa\sqrt{\kappa E_0}}\exp\left[-\frac{\kappa^3}{E_0}\right].
\end{eqnarray}

Now we calculate the corresponding LFA amplitude. In the 1D LFA, the SFA pre-exponential factor ${\cal P}_{2SFA}$ is replaced by
\be
{\cal P}_{LFA}^{(1D)}(t_2)=\frac{E_0t_2}{E_0t_2-i\kappa}{\cal P}_{2SFA},
\ee
see \cite{Krajewska_2010}, and we consequently arrive at
\begin{eqnarray}
m_{LFA}^{(1D)}= \int_{t_{1,s}} dt_2 \; \frac{2\pi}{\kappa}{\cal P}_{LFA}(t_2)  \exp\left[-\frac{\kappa^3}{E_0}-\frac{E_0\kappa}{2} \left(t_2 - i \frac{\kappa}{E_0}\right)^2\right].\nonumber\\
\end{eqnarray}
The latter is calculated analytically in the same way as that of Eq.~(\ref{m2sfa}),
yielding
\begin{eqnarray}
m_{LFA}^{(1D)}&=&  \int dt_2 \; \frac{\pi}{\kappa}{\cal P}_{LFA}(t_2)  \exp\left[-\frac{\kappa^3}{E_0}-\frac{E_0\kappa}{2} \left(t_2 - i \frac{\kappa}{E_0}\right)^2\right]\nonumber\\
&&=\left(1-\sqrt{\frac{\pi}{2}\frac{\kappa^3}{E_0}}\right)m_{2SFA}^{(1D)}.
\end{eqnarray}
Thus, we derive the LFA factor in the 1D case:
\begin{eqnarray}
 {\cal T}_{LFA}^{(1D)}=1-\sqrt{\frac{\pi}{2}\frac{\kappa^3}{E_0}},
\end{eqnarray}
which corrects the SFA recollision amplitude as it incorporates the exact scattering amplitude in the SFA recollision matrix element.

\subsection{3D LFA factor}

In the  LFA for a 3D system with a short-range potential, the SFA pre-exponential factor ${\cal P}_{2SFA}$ is replaced by
\be
{\cal P}_{LFA}=\frac{-i\kappa}{E_0t_2-i\kappa}{\cal P}_{2SFA},
\ee
see \cite{Krajewska_2010}. A calculation similar to 1D gives the following expression for the LFA recollision amplitude in the 3D case:
\begin{eqnarray}
m_{LFA}^{(3D)}&=&\sqrt{\frac{\pi}{2}\frac{\kappa^3}{E_0}}m^{(3D)}_{2SFA}.
\end{eqnarray}
Thus providing the LFA correction factor in the 3D case:
\begin{eqnarray}
 {\cal T}_{LFA}^{(3D)}= \sqrt{\frac{\pi}{2}\frac{\kappa^3}{E_0}}.
\end{eqnarray}

\section{3D wavefunction in momentum space}\label{app_3D_momentum}

The SFA amplitudes in three dimensions contain additional integrations on the transversal coordinate and momentum. We work under the assumption that the final momentum lays in the polarization axis $p_y=p_z=0$, justified by the fact that ionization occurs primarily in the direction of the driving laser field.
Thus, as before, we have the momentum amplitude defined by the integral
\begin{align}
m^{(3D)}_1(p_x,t)&= \int^{t}_{t_i} \,dt_1 \int^{\infty}_{-\infty} dx_1\,\tilde{m}^{(3D)}_1 (p_x, x_1,t_1),
\intertext{
where}
\tilde{m}^{(3D)}_{1}(p_x,x,t)&= \int \int dy_1 \,dz_1 \;\tilde{m}^{(1D)}_{1}(p_x,x_1,t)\frac{\exp\left(-\kappa r_1+\kappa\sqrt{x_1^2}\right)}{(2\pi)^{3/2} \; r_1}\label{m3D1}\nonumber\\
& =2\pi\int d\rho_1  \;\rho_1\tilde{m}^{(1D)}_{1}(p_x,x_1,t)\frac{\exp\left(-\kappa r_1+\kappa\sqrt{x_1^2}\right)}{(2\pi)^{3/2} \; r_1}\nonumber\\
&=\frac{1}{\sqrt{2\pi}\kappa}\tilde{m}^{(1D)}_{1}(p_x,x_1,t)
\end{align}
with $r_1=\sqrt{x_1^2+y_1^2+z_1^2}=\sqrt{x_1^2+\rho_1^2}$.
Thus, the additional integral in Eq.~(\ref{m3D1})
yields an extra factor $c_1=1/\sqrt{2\pi\kappa^2}$ in the first-order amplitude:
\begin{equation}
 \tilde{m}^{(3D)}_{1}(p_x, x_1,t_1)=c_1 \tilde{m}^{(1D)}_{1}(p_x, x_1,t_1).
\end{equation}

Analogously the second-order amplitude in 3D,
\be
m^{(3D)}_2(p_x, t) = \int^{t}_{t_i} dt_1 \int^{t}_{t_1}dt_2 \, \int dx_1 \; \;\tilde{m}^{(3D)}_2(p_x, x_1, t_1,t_2),
\ee
 can be expressed via the corresponding 1D amplitude
 \begin{widetext}
\begin{eqnarray}
\tilde{m}^{(3D)}_{2}(p_x,x_1,t_1,t_2) =- \, \int \, dy_1 \, dz_1 \, dk_y \, dk_z \; \, \tilde{m}^{(1D)}_{2}(p_x,x_1,t_1,t_2) \, \frac{\exp[-i k_y y_1-i k_z z_1-\frac{i}{2} (k_y^2+k_z^2)(t_2-t_1)-\kappa r_1+\kappa \sqrt{x_1^2}]}{\sqrt{(2\pi)^3}\,r_1\kappa^2}
\label{3D_int}
\end{eqnarray}
\end{widetext}

 After the application of a 4D  SPA over $y_1$, $z_1$, $k_y$ and $k_z$ integrations, we obtain the 3D amplitude
\begin{equation}
 \tilde{m}^{(3D)}_{2}(p_x, x_1,t_1,t_2)=c_1 \,c_2(x_1,t_1,t_2) \;\tilde{m}^{(1D)}_{2}(p_x, x_1,t_1,t_2),
\end{equation}
 with the following correction factor to the 1D case:
\begin{equation}
   c_2(x_1,t_1,t_2)=-\frac{1}{i(t_2-t_1)\kappa^2+|x_1|\kappa}.\label{c2xtt}
\end{equation}
In the consequent $x_1$-integration in the second-order amplitude a typical value of $|x_s|\sim 1/\kappa$ in $c_2(x_1)$ is assumed \cite{Suarez_2015}, after which the integration is carried out analytically. This choice is justified because $x_s$ is the $x_1$-saddle point of the product of 1D bound state wavefunction $\sim\exp[-\kappa|x_1|]$ with the interaction Hamiltonian $H_i\sim E(t)x_1$.

Thus, the total amplitude (in the $p_x$ plane) in three dimensions  can be calculated  from the amplitude in one dimension, using Eqs.~(\ref{m1xt}), (\ref{m2xt}):
\begin{eqnarray}
m^{(3D)}_{1}(p_x,t)&=&\int^{t}_{t_i} dt_1\,  \tilde{m}_1^{(3D)}(p_x,0,t_1) {\cal I}_1(t_1),\label{mm1_3D}\\
m^{(3D)}_{2}(p_x,t)&=& \int^t_{t_i} dt_1 \int^{t}_{t_1}dt_2 \tilde{m}_2^{(3D)}(p_x,0,t_1,t_2)\, {\cal I}_2(t_1,t_2)\label{mm2_3D}.
\end{eqnarray}

\section{3D wavefunction in coordinate space}\label{app_3D_coordinate}

 In the 3D case we use the wavefunction in a mixed representation $\Psi_i(x,p_y,p_z,t)$ to derive the Wigner trajectory, choosing the most probable values for the transverse momentum $p_y=p_z=0$:
\begin{eqnarray}
\Psi_i(x,p_y,p_z,t)|_{p_y=p_z=0}&=& \int^{\infty}_{-\infty} dp_x m_{1}^{(3D)}(p_x,t) \Psi_{p_x}(x,t) \\
\Psi_r(x,p_y ,p_z ,t)|_{p_y=p_z=0}&=&\int^{\infty}_{-\infty} dp_x\, m_{2}^{(3D)}(p_x,t) \Psi_{p_x}(x,t).
\end{eqnarray}
With the assumption $p_y=p_z=0$, the spreading of the tunneling wave packet in the continuum, after leaving the tunneling barrier, is neglected. Meanwhile, the spreading during the tunneling inside the barrier has been fully accounted for via the intermediate transverse momentum $k_y,k_z$-integration in Eq.~(\ref{3D_int}). As the tunneling time delay is formed during the tunneling, the latter is relevant for the tunneling time delay, while the former has no effect on the tunneling time delay, and its neglect is thus justified.
The calculations similar to the 1D case provide:
\begin{eqnarray}
 \Psi_i(x,p_y,p_z,t)|_{p_y=p_z=0}&=&\int^{t}_{t_i} dt_1\,  \tilde{m}_1^{(3D)}(\tilde{p}_{s1},0,t_1) {\cal I}_1(t_1)\nonumber\\
 &\times& \sqrt{\frac{2\pi}{i(t-t_1)}} \Psi_{\tilde{p}_{s1}}(x,t),  \label{mmm1}\\
 \Psi_r(x,p_y,p_z,t)|_{p_y=p_z=0}&=&\int^t_{t_i} dt_1 \int^{t}_{t_1}dt_2 \,\tilde{m}_2^{(3D)}(\tilde{p}_{s2},0,t_1,t_2)\, {\cal I}_2(t_1,t_2)\nonumber\\
 &&\times \sqrt{\frac{2\pi}{i(t-t_2)}} \Psi_{\tilde{p}_{s2}}(x,t),\label{mmm2}
\end{eqnarray}
Further, the probability distribution at the exit, $x=x_e$, is calculated via the wavefunction $|\Psi(x_e,p_y=0,p_z=0,t)|^2$.

\section{Calculation of the integral along the vertical contour}
 \label{A1}

 The $V$-SFA momentum amplitude in one dimension in a constant field reads after the two coordinate integrations
\begin{eqnarray}
m_{2SFA}^{(1D)}&=&\int_{-\infty}^{\infty} dt_1 \int_{t_1}^{\infty} dt_2 \int dk\;{\cal P}_{2SFA}\\&& \exp\left[-i\frac{k^2}{2}\left(t_2-t_1\right)-ik\left(\alpha(t_2)-\alpha(t_1)\right)+i\beta(t_1)+i\frac{\kappa^2}{2}t_1\right],\nonumber
\end{eqnarray}
 where $p=0$ was used. We carry out firstly the integration over the intermediate momentum  $k$ by SPA and arrive at
\begin{eqnarray}
m_{2SFA}^{(1D)}&=&\int_{-\infty}^{\infty} dt_1 \int_{t_1}^{\infty} dt_2\; {\cal P}_{2SFA}\; \frac{\sqrt{2\pi}}{\sqrt{ i(t_2-t_1)}}\\&& \exp\left[-i\frac{k_s^2}{2}\left(t_2-t_1\right)-ik_s\left(\alpha(t_2)-\alpha(t_1)\right)+i\beta(t_1)+i\frac{\kappa^2}{2}t_1\right]\nonumber
\end{eqnarray}
with $k_s=-(\alpha(t_2)-\alpha(t_1))/(t_2-t_1)$. The following $t_2$-integral consitsts of a vertical contour from $t_1$ to $t_{2,s}$ and a horizontal contour from $t_{2,s}$ to $\infty$. In this section we want to estimate the first which has its dominant contribution in the region around $t_2=t_1$. We therefore expand the integrand in $t_2$ around $t_1$:
 \begin{eqnarray}
m_{2SFA}^{(1D)}&=-&\int_{-\infty}^{\infty} dt_1 \int_{t_1}^{\infty} dt_2\; \frac{\kappa^{5/2}}{2\pi\sqrt{ i(t_2-t_1)}}\\&& \exp\left[\frac{i}{2} E_0^2 (t_2 - t_1) t_1^2 + \frac{i}{6} (E_0^2 t_1^3 + 3 \kappa^2t_1)\right]\nonumber
\end{eqnarray}
and integrate analytically
 \begin{eqnarray}
m_{2SFA}^{(1D)}&=&\int_{-\infty}^{\infty} dt_1\; \frac{\kappa^{5/2}}{\sqrt{2\pi}E_0t_1}\exp\left[\frac{i}{6} (E_0^2 t_1^3 + 3 \kappa^2 t_1)\right].
\end{eqnarray}
The final integral is again evaluated via SPA, where the exponent is expanded quadratically at the saddle point $t_{1,s}=i \kappa/E_0$ and the latter is  inserted into the  prexponential. With these approximations we derive the direct ionzation amplitude
\begin{eqnarray}
m_{2SFA}^{(1D)}=\frac{i\kappa}{\sqrt{E_0}}\exp\left[-\frac{\kappa^3}{3E_0}\right].
\end{eqnarray}

\section{The numerical time integrations}
 \label{A2}
The derivation of the ionization amplitudes leads to time integrals, that are calculated numerically. We consider intermediate observation times $t$, when the electron is close to the tunnel exit, and
%For intermediate observation times $t$, i.e. in this paper when the electron is close to the tunnel exit,
the structure of the integrands is the following
\begin{eqnarray}
\tilde{m}=\int^t_{t_a} dt' \frac{\exp[f(t')]}{\sqrt{t-t'}},
\end{eqnarray}
where the function $f(t')$ has a singularity at $t'=t$. To handle the integration at the singularity, we single out the singular part of $f(t')$ as $f(t')=-if_{-1}/(t-t')+{\cal O}(1)$, with a constant coefficient $f_{-1}>0$, and rewrite the integral
\begin{eqnarray}
\tilde{m}=\int^t_{t_a} dt' \left(\frac{\exp\left[f(t')\right]}{\sqrt{t-t'}}-\frac{\exp\left[\frac{-if_{-1}}{t-t'}\right]}{\sqrt{t-t'}}\right)+\int^t_{t_a} dt' \frac{\exp\left[\frac{-if_{-1}}{t-t'}\right]}{\sqrt{t-t'}}\nonumber\\
\end{eqnarray}
Now it is possible to perform the first integration numerically, since the singularity is omitted, and second integral to calculate analytically, which yields
\begin{eqnarray}
&&\tilde{m}=\int^t_{t_a} dt' \left(\frac{\exp\left[f(t')\right]}{\sqrt{t-t'}}-\frac{\exp\left[\frac{-if_{-1}}{t-t'}\right]}{\sqrt{t-t'}}\right)-(1 + i) \sqrt{f_{-1}} \sqrt{2 \pi}\\
&&+(1+i) \sqrt{2 \pi } \sqrt{f_{-1}} \text{erf}\left((1+i) \sqrt{\frac{f_{-1}}{2( t- t_a)}}\right)+2 \sqrt{t-t_a} \exp\left[\frac{i f_{-1}}{t_a-t}\right],\nonumber
\end{eqnarray}

For asymptotic observation times $t\rightarrow \infty$, we approximate the integrand function, expanding it near the final time $t_f$:
\begin{eqnarray}
f(t')&\approx& f(t_f)+(t'-t_f) f'(t_f) \nonumber\\
&\approx &f(t_f)\exp\left\{\ln \left[1+ \frac{(t'-t_f) f'(t_f)}{f(t_f)}\right]\right\}\nonumber\\
&\approx&  f(t_f)\exp\left [\frac{(t'-t_f) f'(t_f)}{f(t_f)}\right ],
\end{eqnarray}
and calculate the integrals as follows
\begin{eqnarray}
\tilde{m}&=&\int^\infty_{t_a} dt'f(t')\nonumber\\
&\approx&\int^\infty_{t_a} dt' f(t_f) \exp\left[\frac{(t'-t_f) f'(t_f)}{f(t_f)}\right]\nonumber\\
&\approx& -\frac{f(t_f)^2}{f'(t_f)},
\end{eqnarray}
where $t_f$ is a time after the turn-off of the laser pulse.

\bibliography{strong_fields_bibliography3}

%apsrev4-2.bst 2019-01-14 (MD) hand-edited version of apsrev4-1.bst
%Control: key (0)
%Control: author (8) initials jnrlst
%Control: editor formatted (1) identically to author
%Control: production of article title (0) allowed
%Control: page (0) single
%Control: year (1) truncated
%Control: production of eprint (0) enabled
\begin{thebibliography}{79}%
\makeatletter
\providecommand \@ifxundefined [1]{%
 \@ifx{#1\undefined}
}%
\providecommand \@ifnum [1]{%
 \ifnum #1\expandafter \@firstoftwo
 \else \expandafter \@secondoftwo
 \fi
}%
\providecommand \@ifx [1]{%
 \ifx #1\expandafter \@firstoftwo
 \else \expandafter \@secondoftwo
 \fi
}%
\providecommand \natexlab [1]{#1}%
\providecommand \enquote  [1]{``#1''}%
\providecommand \bibnamefont  [1]{#1}%
\providecommand \bibfnamefont [1]{#1}%
\providecommand \citenamefont [1]{#1}%
\providecommand \href@noop [0]{\@secondoftwo}%
\providecommand \href [0]{\begingroup \@sanitize@url \@href}%
\providecommand \@href[1]{\@@startlink{#1}\@@href}%
\providecommand \@@href[1]{\endgroup#1\@@endlink}%
\providecommand \@sanitize@url [0]{\catcode `\\12\catcode `\$12\catcode
  `\&12\catcode `\#12\catcode `\^12\catcode `\_12\catcode `\%12\relax}%
\providecommand \@@startlink[1]{}%
\providecommand \@@endlink[0]{}%
\providecommand \url  [0]{\begingroup\@sanitize@url \@url }%
\providecommand \@url [1]{\endgroup\@href {#1}{\urlprefix }}%
\providecommand \urlprefix  [0]{URL }%
\providecommand \Eprint [0]{\href }%
\providecommand \doibase [0]{https://doi.org/}%
\providecommand \selectlanguage [0]{\@gobble}%
\providecommand \bibinfo  [0]{\@secondoftwo}%
\providecommand \bibfield  [0]{\@secondoftwo}%
\providecommand \translation [1]{[#1]}%
\providecommand \BibitemOpen [0]{}%
\providecommand \bibitemStop [0]{}%
\providecommand \bibitemNoStop [0]{.\EOS\space}%
\providecommand \EOS [0]{\spacefactor3000\relax}%
\providecommand \BibitemShut  [1]{\csname bibitem#1\endcsname}%
\let\auto@bib@innerbib\@empty
%</preamble>
\bibitem [{\citenamefont {Corkum}\ and\ \citenamefont
  {Krausz}(2007)}]{Corkum_2007}%
  \BibitemOpen
  \bibfield  {author} {\bibinfo {author} {\bibfnamefont {P.~B.}\ \bibnamefont
  {Corkum}}\ and\ \bibinfo {author} {\bibfnamefont {F.}~\bibnamefont
  {Krausz}},\ }\bibfield  {title} {\bibinfo {title} {Attosecond science},\
  }\href@noop {} {\bibfield  {journal} {\bibinfo  {journal} {Nature Phys.}\
  }\textbf {\bibinfo {volume} {3}},\ \bibinfo {pages} {381} (\bibinfo {year}
  {2007})}\BibitemShut {NoStop}%
\bibitem [{\citenamefont {Krausz}\ and\ \citenamefont
  {Ivanov}(2009)}]{Krausz_2009}%
  \BibitemOpen
  \bibfield  {author} {\bibinfo {author} {\bibfnamefont {F.}~\bibnamefont
  {Krausz}}\ and\ \bibinfo {author} {\bibfnamefont {M.}~\bibnamefont
  {Ivanov}},\ }\bibfield  {title} {\bibinfo {title} {{Attosecond physics}},\
  }\href@noop {} {\bibfield  {journal} {\bibinfo  {journal} {Rev. Mod. Phys.}\
  }\textbf {\bibinfo {volume} {81}},\ \bibinfo {pages} {163} (\bibinfo {year}
  {2009})}\BibitemShut {NoStop}%
\bibitem [{\citenamefont {Maharjan}\ \emph {et~al.}(2005)\citenamefont
  {Maharjan}, \citenamefont {Alnaser}, \citenamefont {Tong}, \citenamefont
  {Ulrich}, \citenamefont {Ranitovic}, \citenamefont {Ghimire}, \citenamefont
  {Chang}, \citenamefont {Litvinyuk},\ and\ \citenamefont
  {Cocke}}]{Maharjan_2005}%
  \BibitemOpen
  \bibfield  {author} {\bibinfo {author} {\bibfnamefont {C.~M.}\ \bibnamefont
  {Maharjan}}, \bibinfo {author} {\bibfnamefont {A.~S.}\ \bibnamefont
  {Alnaser}}, \bibinfo {author} {\bibfnamefont {X.~M.}\ \bibnamefont {Tong}},
  \bibinfo {author} {\bibfnamefont {B.}~\bibnamefont {Ulrich}}, \bibinfo
  {author} {\bibfnamefont {P.}~\bibnamefont {Ranitovic}}, \bibinfo {author}
  {\bibfnamefont {S.}~\bibnamefont {Ghimire}}, \bibinfo {author} {\bibfnamefont
  {Z.}~\bibnamefont {Chang}}, \bibinfo {author} {\bibfnamefont {I.~V.}\
  \bibnamefont {Litvinyuk}},\ and\ \bibinfo {author} {\bibfnamefont {C.~L.}\
  \bibnamefont {Cocke}},\ }\bibfield  {title} {\bibinfo {title} {Momentum
  imaging of doubly charged ions of ne and ar in the sequential ionization
  region},\ }\href {https://doi.org/10.1103/PhysRevA.72.041403} {\bibfield
  {journal} {\bibinfo  {journal} {Phys. Rev. A}\ }\textbf {\bibinfo {volume}
  {72}},\ \bibinfo {pages} {041403(R)} (\bibinfo {year} {2005})}\BibitemShut
  {NoStop}%
\bibitem [{\citenamefont {Eckle}\ \emph
  {et~al.}(2008{\natexlab{a}})\citenamefont {Eckle}, \citenamefont {Smolarski},
  \citenamefont {Schlup}, \citenamefont {Biegert}, \citenamefont {Staudte},
  \citenamefont {Sch{\"{o}}ffler}, \citenamefont {Muller}, \citenamefont
  {D{\"{o}}rner},\ and\ \citenamefont {Keller}}]{Eckle_2008a}%
  \BibitemOpen
  \bibfield  {author} {\bibinfo {author} {\bibfnamefont {P.}~\bibnamefont
  {Eckle}}, \bibinfo {author} {\bibfnamefont {M.}~\bibnamefont {Smolarski}},
  \bibinfo {author} {\bibfnamefont {F.}~\bibnamefont {Schlup}}, \bibinfo
  {author} {\bibfnamefont {J.}~\bibnamefont {Biegert}}, \bibinfo {author}
  {\bibfnamefont {A.}~\bibnamefont {Staudte}}, \bibinfo {author} {\bibfnamefont
  {M.}~\bibnamefont {Sch{\"{o}}ffler}}, \bibinfo {author} {\bibfnamefont
  {H.~G.}\ \bibnamefont {Muller}}, \bibinfo {author} {\bibfnamefont
  {R.}~\bibnamefont {D{\"{o}}rner}},\ and\ \bibinfo {author} {\bibfnamefont
  {U.}~\bibnamefont {Keller}},\ }\bibfield  {title} {\bibinfo {title}
  {{Attosecond angular streaking}},\ }\href@noop {} {\bibfield  {journal}
  {\bibinfo  {journal} {Nature Phys.}\ }\textbf {\bibinfo {volume} {4}},\
  \bibinfo {pages} {565} (\bibinfo {year} {2008}{\natexlab{a}})}\BibitemShut
  {NoStop}%
\bibitem [{\citenamefont {Eckle}\ \emph
  {et~al.}(2008{\natexlab{b}})\citenamefont {Eckle}, \citenamefont {Pfeiffer},
  \citenamefont {Cirelli}, \citenamefont {Staudte}, \citenamefont
  {D{\"{o}}rner}, \citenamefont {Muller}, \citenamefont {B{\"{u}}ttiker},\ and\
  \citenamefont {Keller}}]{Eckle_2008b}%
  \BibitemOpen
  \bibfield  {author} {\bibinfo {author} {\bibfnamefont {P.}~\bibnamefont
  {Eckle}}, \bibinfo {author} {\bibfnamefont {A.~N.}\ \bibnamefont {Pfeiffer}},
  \bibinfo {author} {\bibfnamefont {C.}~\bibnamefont {Cirelli}}, \bibinfo
  {author} {\bibfnamefont {A.}~\bibnamefont {Staudte}}, \bibinfo {author}
  {\bibfnamefont {R.}~\bibnamefont {D{\"{o}}rner}}, \bibinfo {author}
  {\bibfnamefont {H.~G.}\ \bibnamefont {Muller}}, \bibinfo {author}
  {\bibfnamefont {M.}~\bibnamefont {B{\"{u}}ttiker}},\ and\ \bibinfo {author}
  {\bibfnamefont {U.}~\bibnamefont {Keller}},\ }\bibfield  {title} {\bibinfo
  {title} {{Attosecond Ionization and Tunneling Delay Time Measurements in
  Helium}},\ }\href@noop {} {\bibfield  {journal} {\bibinfo  {journal}
  {Science}\ }\textbf {\bibinfo {volume} {322}},\ \bibinfo {pages} {1525}
  (\bibinfo {year} {2008}{\natexlab{b}})}\BibitemShut {NoStop}%
\bibitem [{\citenamefont {Pfeiffer}\ \emph {et~al.}(2012)\citenamefont
  {Pfeiffer}, \citenamefont {Cirelli}, \citenamefont {Smolarski}, \citenamefont
  {Dimitrovski}, \citenamefont {Abu-samha}, \citenamefont {Madsen},\ and\
  \citenamefont {Keller}}]{Pfeiffer_2012}%
  \BibitemOpen
  \bibfield  {author} {\bibinfo {author} {\bibfnamefont {A.~N.}\ \bibnamefont
  {Pfeiffer}}, \bibinfo {author} {\bibfnamefont {C.}~\bibnamefont {Cirelli}},
  \bibinfo {author} {\bibfnamefont {M.}~\bibnamefont {Smolarski}}, \bibinfo
  {author} {\bibfnamefont {D.}~\bibnamefont {Dimitrovski}}, \bibinfo {author}
  {\bibfnamefont {M.}~\bibnamefont {Abu-samha}}, \bibinfo {author}
  {\bibfnamefont {L.~B.}\ \bibnamefont {Madsen}},\ and\ \bibinfo {author}
  {\bibfnamefont {U.}~\bibnamefont {Keller}},\ }\bibfield  {title} {\bibinfo
  {title} {{Attoclock reveals natural coordinates of the laser-induced
  tunnelling current flow in atoms}},\ }\href@noop {} {\bibfield  {journal}
  {\bibinfo  {journal} {Nature Phys.}\ }\textbf {\bibinfo {volume} {8}},\
  \bibinfo {pages} {76} (\bibinfo {year} {2012})}\BibitemShut {NoStop}%
\bibitem [{\citenamefont {Landsman}\ \emph {et~al.}(2014)\citenamefont
  {Landsman}, \citenamefont {Weger}, \citenamefont {Maurer}, \citenamefont
  {Boge}, \citenamefont {Ludwig}, \citenamefont {Heuser}, \citenamefont
  {Cirelli}, \citenamefont {Gallmann},\ and\ \citenamefont
  {Keller}}]{Landsman_2014o}%
  \BibitemOpen
  \bibfield  {author} {\bibinfo {author} {\bibfnamefont {A.~S.}\ \bibnamefont
  {Landsman}}, \bibinfo {author} {\bibfnamefont {M.}~\bibnamefont {Weger}},
  \bibinfo {author} {\bibfnamefont {J.}~\bibnamefont {Maurer}}, \bibinfo
  {author} {\bibfnamefont {R.}~\bibnamefont {Boge}}, \bibinfo {author}
  {\bibfnamefont {A.}~\bibnamefont {Ludwig}}, \bibinfo {author} {\bibfnamefont
  {S.}~\bibnamefont {Heuser}}, \bibinfo {author} {\bibfnamefont
  {C.}~\bibnamefont {Cirelli}}, \bibinfo {author} {\bibfnamefont
  {L.}~\bibnamefont {Gallmann}},\ and\ \bibinfo {author} {\bibfnamefont
  {U.}~\bibnamefont {Keller}},\ }\bibfield  {title} {\bibinfo {title}
  {{Ultrafast resolution of tunneling delay time}},\ }\href@noop {} {\bibfield
  {journal} {\bibinfo  {journal} {Optica}\ }\textbf {\bibinfo {volume} {1}},\
  \bibinfo {pages} {343} (\bibinfo {year} {2014})}\BibitemShut {NoStop}%
\bibitem [{\citenamefont {Sainadh}\ \emph {et~al.}(2019)\citenamefont
  {Sainadh}, \citenamefont {Xu}, \citenamefont {Wang}, \citenamefont
  {Atia-Tul-Noor}, \citenamefont {Wallace}, \citenamefont {Douguet},
  \citenamefont {Bray}, \citenamefont {Ivanov}, \citenamefont {Bartschat},
  \citenamefont {Kheifets}, \citenamefont {Sang},\ and\ \citenamefont
  {Litvinyuk}}]{Sainadh_2019}%
  \BibitemOpen
  \bibfield  {author} {\bibinfo {author} {\bibfnamefont {U.~S.}\ \bibnamefont
  {Sainadh}}, \bibinfo {author} {\bibfnamefont {H.}~\bibnamefont {Xu}},
  \bibinfo {author} {\bibfnamefont {X.}~\bibnamefont {Wang}}, \bibinfo {author}
  {\bibfnamefont {A.}~\bibnamefont {Atia-Tul-Noor}}, \bibinfo {author}
  {\bibfnamefont {W.~C.}\ \bibnamefont {Wallace}}, \bibinfo {author}
  {\bibfnamefont {N.}~\bibnamefont {Douguet}}, \bibinfo {author} {\bibfnamefont
  {A.}~\bibnamefont {Bray}}, \bibinfo {author} {\bibfnamefont {I.}~\bibnamefont
  {Ivanov}}, \bibinfo {author} {\bibfnamefont {K.}~\bibnamefont {Bartschat}},
  \bibinfo {author} {\bibfnamefont {A.}~\bibnamefont {Kheifets}}, \bibinfo
  {author} {\bibfnamefont {R.~T.}\ \bibnamefont {Sang}},\ and\ \bibinfo
  {author} {\bibfnamefont {I.~V.}\ \bibnamefont {Litvinyuk}},\ }\bibfield
  {title} {\bibinfo {title} {{Attosecond angular streaking and tunnelling time
  in atomic hydrogen}},\ }\href@noop {} {\bibfield  {journal} {\bibinfo
  {journal} {Nature}\ }\textbf {\bibinfo {volume} {568}},\ \bibinfo {pages}
  {75} (\bibinfo {year} {2019})}\BibitemShut {NoStop}%
\bibitem [{\citenamefont {Camus}\ \emph {et~al.}(2017)\citenamefont {Camus},
  \citenamefont {Yakaboylu}, \citenamefont {Fechner}, \citenamefont {Klaiber},
  \citenamefont {Laux}, \citenamefont {Mi}, \citenamefont {Hatsagortsyan},
  \citenamefont {Pfeifer}, \citenamefont {Keitel},\ and\ \citenamefont
  {Moshammer}}]{Camus_2017}%
  \BibitemOpen
  \bibfield  {author} {\bibinfo {author} {\bibfnamefont {N.}~\bibnamefont
  {Camus}}, \bibinfo {author} {\bibfnamefont {E.}~\bibnamefont {Yakaboylu}},
  \bibinfo {author} {\bibfnamefont {L.}~\bibnamefont {Fechner}}, \bibinfo
  {author} {\bibfnamefont {M.}~\bibnamefont {Klaiber}}, \bibinfo {author}
  {\bibfnamefont {M.}~\bibnamefont {Laux}}, \bibinfo {author} {\bibfnamefont
  {Y.}~\bibnamefont {Mi}}, \bibinfo {author} {\bibfnamefont {K.~Z.}\
  \bibnamefont {Hatsagortsyan}}, \bibinfo {author} {\bibfnamefont
  {T.}~\bibnamefont {Pfeifer}}, \bibinfo {author} {\bibfnamefont {C.~H.}\
  \bibnamefont {Keitel}},\ and\ \bibinfo {author} {\bibfnamefont
  {R.}~\bibnamefont {Moshammer}},\ }\bibfield  {title} {\bibinfo {title}
  {{Experimental evidence for Wigner's tunneling time}},\ }\href@noop {}
  {\bibfield  {journal} {\bibinfo  {journal} {Phys. Rev. Lett.}\ }\textbf
  {\bibinfo {volume} {119}},\ \bibinfo {pages} {023201} (\bibinfo {year}
  {2017})}\BibitemShut {NoStop}%
\bibitem [{\citenamefont {Teeny}\ \emph
  {et~al.}(2016{\natexlab{a}})\citenamefont {Teeny}, \citenamefont {Yakaboylu},
  \citenamefont {Bauke},\ and\ \citenamefont {Keitel}}]{Teeny_2016a}%
  \BibitemOpen
  \bibfield  {author} {\bibinfo {author} {\bibfnamefont {N.}~\bibnamefont
  {Teeny}}, \bibinfo {author} {\bibfnamefont {E.}~\bibnamefont {Yakaboylu}},
  \bibinfo {author} {\bibfnamefont {H.}~\bibnamefont {Bauke}},\ and\ \bibinfo
  {author} {\bibfnamefont {C.~H.}\ \bibnamefont {Keitel}},\ }\bibfield  {title}
  {\bibinfo {title} {{Ionization Time and Exit Momentum in Strong-Field Tunnel
  Ionization}},\ }\href@noop {} {\bibfield  {journal} {\bibinfo  {journal}
  {Phys. Rev. Lett.}\ }\textbf {\bibinfo {volume} {116}},\ \bibinfo {pages}
  {063003} (\bibinfo {year} {2016}{\natexlab{a}})}\BibitemShut {NoStop}%
\bibitem [{\citenamefont {Teeny}\ \emph
  {et~al.}(2016{\natexlab{b}})\citenamefont {Teeny}, \citenamefont {Keitel},\
  and\ \citenamefont {Bauke}}]{Teeny_2016b}%
  \BibitemOpen
  \bibfield  {author} {\bibinfo {author} {\bibfnamefont {N.}~\bibnamefont
  {Teeny}}, \bibinfo {author} {\bibfnamefont {C.~H.}\ \bibnamefont {Keitel}},\
  and\ \bibinfo {author} {\bibfnamefont {H.}~\bibnamefont {Bauke}},\ }\bibfield
   {title} {\bibinfo {title} {{Virtual-detector approach to tunnel ionization
  and tunneling times}},\ }\href@noop {} {\bibfield  {journal} {\bibinfo
  {journal} {Phys. Rev. A}\ }\textbf {\bibinfo {volume} {94}},\ \bibinfo
  {pages} {022104} (\bibinfo {year} {2016}{\natexlab{b}})}\BibitemShut
  {NoStop}%
\bibitem [{\citenamefont {Yakaboylu}\ \emph {et~al.}(2013)\citenamefont
  {Yakaboylu}, \citenamefont {Klaiber}, \citenamefont {Bauke}, \citenamefont
  {Hatsagortsyan},\ and\ \citenamefont {Keitel}}]{Yakaboylu_2013}%
  \BibitemOpen
  \bibfield  {author} {\bibinfo {author} {\bibfnamefont {E.}~\bibnamefont
  {Yakaboylu}}, \bibinfo {author} {\bibfnamefont {M.}~\bibnamefont {Klaiber}},
  \bibinfo {author} {\bibfnamefont {H.}~\bibnamefont {Bauke}}, \bibinfo
  {author} {\bibfnamefont {K.~Z.}\ \bibnamefont {Hatsagortsyan}},\ and\
  \bibinfo {author} {\bibfnamefont {C.~H.}\ \bibnamefont {Keitel}},\ }\bibfield
   {title} {\bibinfo {title} {{Relativistic features and time delay of
  laser-induced tunnel ionization}},\ }\href@noop {} {\bibfield  {journal}
  {\bibinfo  {journal} {Phys. Rev. A}\ }\textbf {\bibinfo {volume} {88}},\
  \bibinfo {pages} {063421} (\bibinfo {year} {2013})}\BibitemShut {NoStop}%
\bibitem [{\citenamefont {Han}\ \emph {et~al.}(2019)\citenamefont {Han},
  \citenamefont {Ge}, \citenamefont {Fang}, \citenamefont {Yu}, \citenamefont
  {Guo}, \citenamefont {Ma}, \citenamefont {Deng}, \citenamefont {Gong},\ and\
  \citenamefont {Liu}}]{Han_2019}%
  \BibitemOpen
  \bibfield  {author} {\bibinfo {author} {\bibfnamefont {M.}~\bibnamefont
  {Han}}, \bibinfo {author} {\bibfnamefont {P.}~\bibnamefont {Ge}}, \bibinfo
  {author} {\bibfnamefont {Y.}~\bibnamefont {Fang}}, \bibinfo {author}
  {\bibfnamefont {X.}~\bibnamefont {Yu}}, \bibinfo {author} {\bibfnamefont
  {Z.}~\bibnamefont {Guo}}, \bibinfo {author} {\bibfnamefont {X.}~\bibnamefont
  {Ma}}, \bibinfo {author} {\bibfnamefont {Y.}~\bibnamefont {Deng}}, \bibinfo
  {author} {\bibfnamefont {Q.}~\bibnamefont {Gong}},\ and\ \bibinfo {author}
  {\bibfnamefont {Y.}~\bibnamefont {Liu}},\ }\bibfield  {title} {\bibinfo
  {title} {{Unifying Tunneling Pictures of Strong-Field Ionization with an
  Improved Attoclock}},\ }\href@noop {} {\bibfield  {journal} {\bibinfo
  {journal} {Phys. Rev. Lett.}\ }\textbf {\bibinfo {volume} {123}},\ \bibinfo
  {pages} {073201} (\bibinfo {year} {2019})}\BibitemShut {NoStop}%
\bibitem [{\citenamefont {Orlando}\ \emph
  {et~al.}(2014{\natexlab{a}})\citenamefont {Orlando}, \citenamefont
  {McDonald}, \citenamefont {Protik}, \citenamefont {Vampa},\ and\
  \citenamefont {Brabec}}]{Orlando_2014}%
  \BibitemOpen
  \bibfield  {author} {\bibinfo {author} {\bibfnamefont {G.}~\bibnamefont
  {Orlando}}, \bibinfo {author} {\bibfnamefont {C.~R.}\ \bibnamefont
  {McDonald}}, \bibinfo {author} {\bibfnamefont {N.~H.}\ \bibnamefont
  {Protik}}, \bibinfo {author} {\bibfnamefont {G.}~\bibnamefont {Vampa}},\ and\
  \bibinfo {author} {\bibfnamefont {T.}~\bibnamefont {Brabec}},\ }\bibfield
  {title} {\bibinfo {title} {{Tunnelling time, what does it mean?}},\
  }\href@noop {} {\bibfield  {journal} {\bibinfo  {journal} {J. Phys. B}\
  }\textbf {\bibinfo {volume} {47}},\ \bibinfo {pages} {204002} (\bibinfo
  {year} {2014}{\natexlab{a}})}\BibitemShut {NoStop}%
\bibitem [{\citenamefont {Orlando}\ \emph
  {et~al.}(2014{\natexlab{b}})\citenamefont {Orlando}, \citenamefont
  {McDonald}, \citenamefont {Protik},\ and\ \citenamefont
  {Brabec}}]{Orlando_2014PRA}%
  \BibitemOpen
  \bibfield  {author} {\bibinfo {author} {\bibfnamefont {G.}~\bibnamefont
  {Orlando}}, \bibinfo {author} {\bibfnamefont {C.~R.}\ \bibnamefont
  {McDonald}}, \bibinfo {author} {\bibfnamefont {N.~H.}\ \bibnamefont
  {Protik}},\ and\ \bibinfo {author} {\bibfnamefont {T.}~\bibnamefont
  {Brabec}},\ }\bibfield  {title} {\bibinfo {title} {{Identification of the
  Keldysh time as a lower limit for the tunneling time}},\ }\href@noop {}
  {\bibfield  {journal} {\bibinfo  {journal} {Phys. Rev. A}\ }\textbf {\bibinfo
  {volume} {89}},\ \bibinfo {pages} {014102} (\bibinfo {year}
  {2014}{\natexlab{b}})}\BibitemShut {NoStop}%
\bibitem [{\citenamefont {Lein}(2011)}]{Lein_2011}%
  \BibitemOpen
  \bibfield  {author} {\bibinfo {author} {\bibfnamefont {M.}~\bibnamefont
  {Lein}},\ }\bibfield  {title} {\bibinfo {title} {{Streaking analysis of
  strong-field ionisation}},\ }\href@noop {} {\bibfield  {journal} {\bibinfo
  {journal} {J. Mod. Opt.}\ }\textbf {\bibinfo {volume} {58}},\ \bibinfo
  {pages} {1188} (\bibinfo {year} {2011})}\BibitemShut {NoStop}%
\bibitem [{\citenamefont {Landsman}\ and\ \citenamefont
  {Keller}(2014)}]{Landsman_2014}%
  \BibitemOpen
  \bibfield  {author} {\bibinfo {author} {\bibfnamefont {A.~S.}\ \bibnamefont
  {Landsman}}\ and\ \bibinfo {author} {\bibfnamefont {U.}~\bibnamefont
  {Keller}},\ }\bibfield  {title} {\bibinfo {title} {{Tunnelling time in strong
  field ionisation}},\ }\href@noop {} {\bibfield  {journal} {\bibinfo
  {journal} {J. Phys. B}\ }\textbf {\bibinfo {volume} {47}},\ \bibinfo {pages}
  {204024} (\bibinfo {year} {2014})}\BibitemShut {NoStop}%
\bibitem [{\citenamefont {Torlina}\ \emph {et~al.}(2015)\citenamefont
  {Torlina}, \citenamefont {Morales}, \citenamefont {Kaushal}, \citenamefont
  {Ivanov}, \citenamefont {Kheifets}, \citenamefont {Zielinski}, \citenamefont
  {Scrinzi}, \citenamefont {Muller}, \citenamefont {Sukiasyan}, \citenamefont
  {Ivanov},\ and\ \citenamefont {Smirnova}}]{Torlina_2015}%
  \BibitemOpen
  \bibfield  {author} {\bibinfo {author} {\bibfnamefont {L.}~\bibnamefont
  {Torlina}}, \bibinfo {author} {\bibfnamefont {F.}~\bibnamefont {Morales}},
  \bibinfo {author} {\bibfnamefont {J.}~\bibnamefont {Kaushal}}, \bibinfo
  {author} {\bibfnamefont {I.}~\bibnamefont {Ivanov}}, \bibinfo {author}
  {\bibfnamefont {A.}~\bibnamefont {Kheifets}}, \bibinfo {author}
  {\bibfnamefont {A.}~\bibnamefont {Zielinski}}, \bibinfo {author}
  {\bibfnamefont {A.}~\bibnamefont {Scrinzi}}, \bibinfo {author} {\bibfnamefont
  {H.~G.}\ \bibnamefont {Muller}}, \bibinfo {author} {\bibfnamefont
  {S.}~\bibnamefont {Sukiasyan}}, \bibinfo {author} {\bibfnamefont
  {M.}~\bibnamefont {Ivanov}},\ and\ \bibinfo {author} {\bibfnamefont
  {O.}~\bibnamefont {Smirnova}},\ }\bibfield  {title} {\bibinfo {title}
  {{Interpreting attoclock measurements of tunnelling times}},\ }\href@noop {}
  {\bibfield  {journal} {\bibinfo  {journal} {Nat. Phys.}\ }\textbf {\bibinfo
  {volume} {11}},\ \bibinfo {pages} {503} (\bibinfo {year} {2015})}\BibitemShut
  {NoStop}%
\bibitem [{\citenamefont {Ni}\ \emph {et~al.}(2018{\natexlab{a}})\citenamefont
  {Ni}, \citenamefont {Saalmann},\ and\ \citenamefont {Rost}}]{Ni_2018b}%
  \BibitemOpen
  \bibfield  {author} {\bibinfo {author} {\bibfnamefont {H.}~\bibnamefont
  {Ni}}, \bibinfo {author} {\bibfnamefont {U.}~\bibnamefont {Saalmann}},\ and\
  \bibinfo {author} {\bibfnamefont {J.-M.}\ \bibnamefont {Rost}},\ }\bibfield
  {title} {\bibinfo {title} {{Tunneling exit characteristics from classical
  backpropagation of an ionized electron wave packet}},\ }\href@noop {}
  {\bibfield  {journal} {\bibinfo  {journal} {Phys. Rev. A}\ }\textbf {\bibinfo
  {volume} {97}},\ \bibinfo {pages} {013426} (\bibinfo {year}
  {2018}{\natexlab{a}})}\BibitemShut {NoStop}%
\bibitem [{\citenamefont {Ni}\ \emph {et~al.}(2018{\natexlab{b}})\citenamefont
  {Ni}, \citenamefont {Eicke}, \citenamefont {Ruiz}, \citenamefont {Cai},
  \citenamefont {Oppermann}, \citenamefont {Shvetsov-Shilovski},\ and\
  \citenamefont {Pi}}]{Ni_2018a}%
  \BibitemOpen
  \bibfield  {author} {\bibinfo {author} {\bibfnamefont {H.}~\bibnamefont
  {Ni}}, \bibinfo {author} {\bibfnamefont {N.}~\bibnamefont {Eicke}}, \bibinfo
  {author} {\bibfnamefont {C.}~\bibnamefont {Ruiz}}, \bibinfo {author}
  {\bibfnamefont {J.}~\bibnamefont {Cai}}, \bibinfo {author} {\bibfnamefont
  {F.}~\bibnamefont {Oppermann}}, \bibinfo {author} {\bibfnamefont {N.~I.}\
  \bibnamefont {Shvetsov-Shilovski}},\ and\ \bibinfo {author} {\bibfnamefont
  {L.-W.}\ \bibnamefont {Pi}},\ }\bibfield  {title} {\bibinfo {title}
  {{Tunneling criteria and a nonadiabatic term for strong-field ionization}},\
  }\href@noop {} {\bibfield  {journal} {\bibinfo  {journal} {Phys. Rev. A}\
  }\textbf {\bibinfo {volume} {98}},\ \bibinfo {pages} {013411} (\bibinfo
  {year} {2018}{\natexlab{b}})}\BibitemShut {NoStop}%
\bibitem [{\citenamefont {Ni}\ \emph {et~al.}(2016)\citenamefont {Ni},
  \citenamefont {Saalmann},\ and\ \citenamefont {Rost}}]{Ni_2016}%
  \BibitemOpen
  \bibfield  {author} {\bibinfo {author} {\bibfnamefont {H.}~\bibnamefont
  {Ni}}, \bibinfo {author} {\bibfnamefont {U.}~\bibnamefont {Saalmann}},\ and\
  \bibinfo {author} {\bibfnamefont {J.-M.}\ \bibnamefont {Rost}},\ }\bibfield
  {title} {\bibinfo {title} {{Tunneling Ionization Time Resolved by
  Backpropagation}},\ }\href@noop {} {\bibfield  {journal} {\bibinfo  {journal}
  {Phys. Rev. Lett.}\ }\textbf {\bibinfo {volume} {117}},\ \bibinfo {pages}
  {023002} (\bibinfo {year} {2016})}\BibitemShut {NoStop}%
\bibitem [{\citenamefont {Hofmann}\ \emph {et~al.}(2021)\citenamefont
  {Hofmann}, \citenamefont {Bray}, \citenamefont {Koch}, \citenamefont {Ni},\
  and\ \citenamefont {Shvetsov-Shilovski}}]{Hofmann_2021}%
  \BibitemOpen
  \bibfield  {author} {\bibinfo {author} {\bibfnamefont {C.}~\bibnamefont
  {Hofmann}}, \bibinfo {author} {\bibfnamefont {A.}~\bibnamefont {Bray}},
  \bibinfo {author} {\bibfnamefont {W.}~\bibnamefont {Koch}}, \bibinfo {author}
  {\bibfnamefont {H.}~\bibnamefont {Ni}},\ and\ \bibinfo {author}
  {\bibfnamefont {N.~I.}\ \bibnamefont {Shvetsov-Shilovski}},\ }\bibfield
  {title} {\bibinfo {title} {{Quantum battles in attoscience: tunnelling}},\
  }\href {https://doi.org/10.1140/epjd/s10053-021-00224-2} {\bibfield
  {journal} {\bibinfo  {journal} {The European Physical Journal D}\ }\textbf
  {\bibinfo {volume} {75}},\ \bibinfo {pages} {208} (\bibinfo {year}
  {2021})}\BibitemShut {NoStop}%
\bibitem [{\citenamefont {Landsman}\ and\ \citenamefont
  {Keller}(2015)}]{Landsman_2015}%
  \BibitemOpen
  \bibfield  {author} {\bibinfo {author} {\bibfnamefont {A.~S.}\ \bibnamefont
  {Landsman}}\ and\ \bibinfo {author} {\bibfnamefont {U.}~\bibnamefont
  {Keller}},\ }\bibfield  {title} {\bibinfo {title} {{Attosecond science and
  the tunnelling time problem}},\ }\href@noop {} {\bibfield  {journal}
  {\bibinfo  {journal} {Phys. Rep.}\ }\textbf {\bibinfo {volume} {547}},\
  \bibinfo {pages} {1} (\bibinfo {year} {2015})}\BibitemShut {NoStop}%
\bibitem [{\citenamefont {Zheltikov}(2016)}]{Zheltikov_2016}%
  \BibitemOpen
  \bibfield  {author} {\bibinfo {author} {\bibfnamefont {A.~M.}\ \bibnamefont
  {Zheltikov}},\ }\bibfield  {title} {\bibinfo {title} {Keldysh parameter,
  photoionization adiabaticity, and the tunneling time},\ }\href
  {https://doi.org/10.1103/PhysRevA.94.043412} {\bibfield  {journal} {\bibinfo
  {journal} {Phys. Rev. A}\ }\textbf {\bibinfo {volume} {94}},\ \bibinfo
  {pages} {043412} (\bibinfo {year} {2016})}\BibitemShut {NoStop}%
\bibitem [{\citenamefont {Zimmermann}\ \emph {et~al.}(2016)\citenamefont
  {Zimmermann}, \citenamefont {Mishra}, \citenamefont {Doran}, \citenamefont
  {Gordon},\ and\ \citenamefont {Landsman}}]{Zimmermann_2016}%
  \BibitemOpen
  \bibfield  {author} {\bibinfo {author} {\bibfnamefont {T.}~\bibnamefont
  {Zimmermann}}, \bibinfo {author} {\bibfnamefont {S.}~\bibnamefont {Mishra}},
  \bibinfo {author} {\bibfnamefont {B.~R.}\ \bibnamefont {Doran}}, \bibinfo
  {author} {\bibfnamefont {D.~F.}\ \bibnamefont {Gordon}},\ and\ \bibinfo
  {author} {\bibfnamefont {A.~S.}\ \bibnamefont {Landsman}},\ }\bibfield
  {title} {\bibinfo {title} {{Tunneling Time and Weak Measurement in Strong
  Field Ionization}},\ }\href@noop {} {\bibfield  {journal} {\bibinfo
  {journal} {Phys. Rev. Lett.}\ }\textbf {\bibinfo {volume} {116}},\ \bibinfo
  {pages} {233603} (\bibinfo {year} {2016})}\BibitemShut {NoStop}%
\bibitem [{\citenamefont {Liu}\ \emph {et~al.}(2017)\citenamefont {Liu},
  \citenamefont {Fu}, \citenamefont {Chen}, \citenamefont {L{\"{u}}},
  \citenamefont {Zhao}, \citenamefont {Yuan},\ and\ \citenamefont
  {Zhao}}]{Liu_2017}%
  \BibitemOpen
  \bibfield  {author} {\bibinfo {author} {\bibfnamefont {J.}~\bibnamefont
  {Liu}}, \bibinfo {author} {\bibfnamefont {Y.}~\bibnamefont {Fu}}, \bibinfo
  {author} {\bibfnamefont {W.}~\bibnamefont {Chen}}, \bibinfo {author}
  {\bibfnamefont {Z.}~\bibnamefont {L{\"{u}}}}, \bibinfo {author}
  {\bibfnamefont {J.}~\bibnamefont {Zhao}}, \bibinfo {author} {\bibfnamefont
  {J.}~\bibnamefont {Yuan}},\ and\ \bibinfo {author} {\bibfnamefont
  {Z.}~\bibnamefont {Zhao}},\ }\bibfield  {title} {\bibinfo {title} {{Offset
  angles of photocurrents generated in few-cycle circularly polarized laser
  fields}},\ }\href {https://doi.org/10.1088/1361-6455/aa575b} {\bibfield
  {journal} {\bibinfo  {journal} {J. Phys. B}\ }\textbf {\bibinfo {volume}
  {50}},\ \bibinfo {pages} {55602} (\bibinfo {year} {2017})}\BibitemShut
  {NoStop}%
\bibitem [{\citenamefont {Song}\ \emph {et~al.}(2017)\citenamefont {Song},
  \citenamefont {Yang}, \citenamefont {Guo},\ and\ \citenamefont
  {Li}}]{Song_2017}%
  \BibitemOpen
  \bibfield  {author} {\bibinfo {author} {\bibfnamefont {Y.}~\bibnamefont
  {Song}}, \bibinfo {author} {\bibfnamefont {Y.}~\bibnamefont {Yang}}, \bibinfo
  {author} {\bibfnamefont {F.}~\bibnamefont {Guo}},\ and\ \bibinfo {author}
  {\bibfnamefont {S.}~\bibnamefont {Li}},\ }\bibfield  {title} {\bibinfo
  {title} {{Revisiting the time-dependent ionization process through the
  Bohmian-mechanics method}},\ }\href@noop {} {\bibfield  {journal} {\bibinfo
  {journal} {J. Phys. B}\ }\textbf {\bibinfo {volume} {50}},\ \bibinfo {pages}
  {95003} (\bibinfo {year} {2017})}\BibitemShut {NoStop}%
\bibitem [{\citenamefont {Yuan}\ \emph {et~al.}(2017)\citenamefont {Yuan},
  \citenamefont {Xin}, \citenamefont {Chu},\ and\ \citenamefont
  {Liu}}]{Yuan_2017}%
  \BibitemOpen
  \bibfield  {author} {\bibinfo {author} {\bibfnamefont {M.}~\bibnamefont
  {Yuan}}, \bibinfo {author} {\bibfnamefont {P.}~\bibnamefont {Xin}}, \bibinfo
  {author} {\bibfnamefont {T.}~\bibnamefont {Chu}},\ and\ \bibinfo {author}
  {\bibfnamefont {H.}~\bibnamefont {Liu}},\ }\bibfield  {title} {\bibinfo
  {title} {Exploring tunneling time by instantaneous ionization rate in
  strong-field ionization},\ }\href {https://doi.org/10.1364/OE.25.023493}
  {\bibfield  {journal} {\bibinfo  {journal} {Opt. Express}\ }\textbf {\bibinfo
  {volume} {25}},\ \bibinfo {pages} {23493} (\bibinfo {year}
  {2017})}\BibitemShut {NoStop}%
\bibitem [{\citenamefont {Klaiber}\ \emph {et~al.}(2018)\citenamefont
  {Klaiber}, \citenamefont {Hatsagortsyan},\ and\ \citenamefont
  {Keitel}}]{Klaiber_2018}%
  \BibitemOpen
  \bibfield  {author} {\bibinfo {author} {\bibfnamefont {M.}~\bibnamefont
  {Klaiber}}, \bibinfo {author} {\bibfnamefont {K.~Z.}\ \bibnamefont
  {Hatsagortsyan}},\ and\ \bibinfo {author} {\bibfnamefont {C.~H.}\
  \bibnamefont {Keitel}},\ }\bibfield  {title} {\bibinfo {title}
  {{Under-the-Tunneling-Barrier Recollisions in Strong-Field Ionization}},\
  }\href@noop {} {\bibfield  {journal} {\bibinfo  {journal} {Phys. Rev. Lett.}\
  }\textbf {\bibinfo {volume} {120}},\ \bibinfo {pages} {013201} (\bibinfo
  {year} {2018})}\BibitemShut {NoStop}%
\bibitem [{\citenamefont {Eicke}\ and\ \citenamefont
  {Lein}(2018)}]{Eicke_2018}%
  \BibitemOpen
  \bibfield  {author} {\bibinfo {author} {\bibfnamefont {N.}~\bibnamefont
  {Eicke}}\ and\ \bibinfo {author} {\bibfnamefont {M.}~\bibnamefont {Lein}},\
  }\bibfield  {title} {\bibinfo {title} {{Trajectory-free ionization times in
  strong-field ionization}},\ }\href@noop {} {\bibfield  {journal} {\bibinfo
  {journal} {Phys. Rev. A}\ }\textbf {\bibinfo {volume} {97}},\ \bibinfo
  {pages} {031402(R)} (\bibinfo {year} {2018})}\BibitemShut {NoStop}%
\bibitem [{\citenamefont {Douguet}\ and\ \citenamefont
  {Bartschat}(2018)}]{Douguet_2018}%
  \BibitemOpen
  \bibfield  {author} {\bibinfo {author} {\bibfnamefont {N.}~\bibnamefont
  {Douguet}}\ and\ \bibinfo {author} {\bibfnamefont {K.}~\bibnamefont
  {Bartschat}},\ }\bibfield  {title} {\bibinfo {title} {{Dynamics of Tunneling
  Ionization using Bohmian Mechanics}},\ }\href@noop {} {\bibfield  {journal}
  {\bibinfo  {journal} {Phys. Rev. A}\ }\textbf {\bibinfo {volume} {97}},\
  \bibinfo {pages} {013402} (\bibinfo {year} {2018})}\BibitemShut {NoStop}%
\bibitem [{\citenamefont {Bray}\ \emph {et~al.}(2018)\citenamefont {Bray},
  \citenamefont {Eckart},\ and\ \citenamefont {Kheifets}}]{Bray_2018}%
  \BibitemOpen
  \bibfield  {author} {\bibinfo {author} {\bibfnamefont {A.~W.}\ \bibnamefont
  {Bray}}, \bibinfo {author} {\bibfnamefont {S.}~\bibnamefont {Eckart}},\ and\
  \bibinfo {author} {\bibfnamefont {A.~S.}\ \bibnamefont {Kheifets}},\
  }\bibfield  {title} {\bibinfo {title} {{Keldysh-Rutherford Model for the
  Attoclock}},\ }\href@noop {} {\bibfield  {journal} {\bibinfo  {journal}
  {Phys. Rev. Lett.}\ }\textbf {\bibinfo {volume} {121}},\ \bibinfo {pages}
  {123201} (\bibinfo {year} {2018})}\BibitemShut {NoStop}%
\bibitem [{\citenamefont {Bayta\ifmmode~\mbox{\c{s}}\else \c{s}\fi{}}\ \emph
  {et~al.}(2018)\citenamefont {Bayta\ifmmode~\mbox{\c{s}}\else \c{s}\fi{}},
  \citenamefont {Bojowald},\ and\ \citenamefont {Crowe}}]{Crowe_2018}%
  \BibitemOpen
  \bibfield  {author} {\bibinfo {author} {\bibfnamefont {B.}~\bibnamefont
  {Bayta\ifmmode~\mbox{\c{s}}\else \c{s}\fi{}}}, \bibinfo {author}
  {\bibfnamefont {M.}~\bibnamefont {Bojowald}},\ and\ \bibinfo {author}
  {\bibfnamefont {S.}~\bibnamefont {Crowe}},\ }\bibfield  {title} {\bibinfo
  {title} {Canonical tunneling time in ionization experiments},\ }\href
  {https://doi.org/10.1103/PhysRevA.98.063417} {\bibfield  {journal} {\bibinfo
  {journal} {Phys. Rev. A}\ }\textbf {\bibinfo {volume} {98}},\ \bibinfo
  {pages} {063417} (\bibinfo {year} {2018})}\BibitemShut {NoStop}%
\bibitem [{\citenamefont {Ren}\ \emph {et~al.}(2018)\citenamefont {Ren},
  \citenamefont {Wu}, \citenamefont {Wang},\ and\ \citenamefont
  {Zheng}}]{Ren_2018}%
  \BibitemOpen
  \bibfield  {author} {\bibinfo {author} {\bibfnamefont {X.}~\bibnamefont
  {Ren}}, \bibinfo {author} {\bibfnamefont {Y.}~\bibnamefont {Wu}}, \bibinfo
  {author} {\bibfnamefont {L.}~\bibnamefont {Wang}},\ and\ \bibinfo {author}
  {\bibfnamefont {Y.}~\bibnamefont {Zheng}},\ }\bibfield  {title} {\bibinfo
  {title} {{Entangled trajectories during ionization of an H atom driven by
  n-cycle laser pulse}},\ }\href@noop {} {\bibfield  {journal} {\bibinfo
  {journal} {Phys. Lett. A}\ }\textbf {\bibinfo {volume} {382}},\ \bibinfo
  {pages} {2662} (\bibinfo {year} {2018})}\BibitemShut {NoStop}%
\bibitem [{\citenamefont {Tan}\ \emph {et~al.}(2018)\citenamefont {Tan},
  \citenamefont {Zhou}, \citenamefont {He}, \citenamefont {Chen}, \citenamefont
  {Ke}, \citenamefont {Liang}, \citenamefont {Zhu}, \citenamefont {Li},\ and\
  \citenamefont {Lu}}]{Tan_2018}%
  \BibitemOpen
  \bibfield  {author} {\bibinfo {author} {\bibfnamefont {J.}~\bibnamefont
  {Tan}}, \bibinfo {author} {\bibfnamefont {Y.}~\bibnamefont {Zhou}}, \bibinfo
  {author} {\bibfnamefont {M.}~\bibnamefont {He}}, \bibinfo {author}
  {\bibfnamefont {Y.}~\bibnamefont {Chen}}, \bibinfo {author} {\bibfnamefont
  {Q.}~\bibnamefont {Ke}}, \bibinfo {author} {\bibfnamefont {J.}~\bibnamefont
  {Liang}}, \bibinfo {author} {\bibfnamefont {X.}~\bibnamefont {Zhu}}, \bibinfo
  {author} {\bibfnamefont {M.}~\bibnamefont {Li}},\ and\ \bibinfo {author}
  {\bibfnamefont {P.}~\bibnamefont {Lu}},\ }\bibfield  {title} {\bibinfo
  {title} {{Determination of the Ionization Time Using Attosecond Photoelectron
  Interferometry}},\ }\href@noop {} {\bibfield  {journal} {\bibinfo  {journal}
  {Phys. Rev. Lett.}\ }\textbf {\bibinfo {volume} {121}},\ \bibinfo {pages}
  {253203} (\bibinfo {year} {2018})}\BibitemShut {NoStop}%
\bibitem [{\citenamefont {Sokolovski}\ and\ \citenamefont
  {Akhmatskaya}(2018)}]{Sokolovski_2018}%
  \BibitemOpen
  \bibfield  {author} {\bibinfo {author} {\bibfnamefont {D.}~\bibnamefont
  {Sokolovski}}\ and\ \bibinfo {author} {\bibfnamefont {E.}~\bibnamefont
  {Akhmatskaya}},\ }\bibfield  {title} {\bibinfo {title} {{No time at the end
  of the tunnel}},\ }\href@noop {} {\bibfield  {journal} {\bibinfo  {journal}
  {Commun. Phys.}\ }\textbf {\bibinfo {volume} {1}},\ \bibinfo {pages} {47}
  (\bibinfo {year} {2018})}\BibitemShut {NoStop}%
\bibitem [{\citenamefont {Quan}\ \emph {et~al.}(2019)\citenamefont {Quan},
  \citenamefont {Serov}, \citenamefont {Wei}, \citenamefont {Zhao},
  \citenamefont {Zhou}, \citenamefont {Wang}, \citenamefont {Lai},
  \citenamefont {Kheifets},\ and\ \citenamefont {Liu}}]{Quan_2019}%
  \BibitemOpen
  \bibfield  {author} {\bibinfo {author} {\bibfnamefont {W.}~\bibnamefont
  {Quan}}, \bibinfo {author} {\bibfnamefont {V.~V.}\ \bibnamefont {Serov}},
  \bibinfo {author} {\bibfnamefont {M.~Z.}\ \bibnamefont {Wei}}, \bibinfo
  {author} {\bibfnamefont {M.}~\bibnamefont {Zhao}}, \bibinfo {author}
  {\bibfnamefont {Y.}~\bibnamefont {Zhou}}, \bibinfo {author} {\bibfnamefont
  {Y.~L.}\ \bibnamefont {Wang}}, \bibinfo {author} {\bibfnamefont {X.~Y.}\
  \bibnamefont {Lai}}, \bibinfo {author} {\bibfnamefont {A.~S.}\ \bibnamefont
  {Kheifets}},\ and\ \bibinfo {author} {\bibfnamefont {X.~J.}\ \bibnamefont
  {Liu}},\ }\bibfield  {title} {\bibinfo {title} {{Attosecond Molecular Angular
  Streaking with All-Ionic Fragments Detection}},\ }\href@noop {} {\bibfield
  {journal} {\bibinfo  {journal} {Phys. Rev. Lett.}\ }\textbf {\bibinfo
  {volume} {123}},\ \bibinfo {pages} {223204} (\bibinfo {year}
  {2019})}\BibitemShut {NoStop}%
\bibitem [{\citenamefont {Douguet}\ and\ \citenamefont
  {Bartschat}(2019)}]{Douguet_2019}%
  \BibitemOpen
  \bibfield  {author} {\bibinfo {author} {\bibfnamefont {N.}~\bibnamefont
  {Douguet}}\ and\ \bibinfo {author} {\bibfnamefont {K.}~\bibnamefont
  {Bartschat}},\ }\bibfield  {title} {\bibinfo {title} {{Attoclock setup with
  negative ions: A possibility for experimental validation}},\ }\href@noop {}
  {\bibfield  {journal} {\bibinfo  {journal} {Phys. Rev. A}\ }\textbf {\bibinfo
  {volume} {99}},\ \bibinfo {pages} {023417} (\bibinfo {year}
  {2019})}\BibitemShut {NoStop}%
\bibitem [{\citenamefont {Hofmann}\ \emph {et~al.}(2019)\citenamefont
  {Hofmann}, \citenamefont {Landsman},\ and\ \citenamefont
  {Keller}}]{Hofmann_2019}%
  \BibitemOpen
  \bibfield  {author} {\bibinfo {author} {\bibfnamefont {C.}~\bibnamefont
  {Hofmann}}, \bibinfo {author} {\bibfnamefont {A.~S.}\ \bibnamefont
  {Landsman}},\ and\ \bibinfo {author} {\bibfnamefont {U.}~\bibnamefont
  {Keller}},\ }\bibfield  {title} {\bibinfo {title} {{Attoclock revisited on
  electron tunnelling time}},\ }\href@noop {} {\bibfield  {journal} {\bibinfo
  {journal} {J. Mod. Opt.}\ }\textbf {\bibinfo {volume} {66}},\ \bibinfo
  {pages} {1052} (\bibinfo {year} {2019})}\BibitemShut {NoStop}%
\bibitem [{\citenamefont {Serov}\ \emph {et~al.}(2019)\citenamefont {Serov},
  \citenamefont {Bray},\ and\ \citenamefont {Kheifets}}]{Serov_2019}%
  \BibitemOpen
  \bibfield  {author} {\bibinfo {author} {\bibfnamefont {V.~V.}\ \bibnamefont
  {Serov}}, \bibinfo {author} {\bibfnamefont {A.~W.}\ \bibnamefont {Bray}},\
  and\ \bibinfo {author} {\bibfnamefont {A.~S.}\ \bibnamefont {Kheifets}},\
  }\bibfield  {title} {\bibinfo {title} {{Numerical attoclock on atomic and
  molecular hydrogen}},\ }\href@noop {} {\bibfield  {journal} {\bibinfo
  {journal} {Phys. Rev. A}\ }\textbf {\bibinfo {volume} {99}},\ \bibinfo
  {pages} {063428} (\bibinfo {year} {2019})}\BibitemShut {NoStop}%
\bibitem [{\citenamefont {Wang}\ \emph {et~al.}(2019)\citenamefont {Wang},
  \citenamefont {Zhang}, \citenamefont {Li}, \citenamefont {Xu}, \citenamefont
  {Cao}, \citenamefont {Zhou}, \citenamefont {Cao},\ and\ \citenamefont
  {Lu}}]{Wang_2019}%
  \BibitemOpen
  \bibfield  {author} {\bibinfo {author} {\bibfnamefont {R.}~\bibnamefont
  {Wang}}, \bibinfo {author} {\bibfnamefont {Q.}~\bibnamefont {Zhang}},
  \bibinfo {author} {\bibfnamefont {D.}~\bibnamefont {Li}}, \bibinfo {author}
  {\bibfnamefont {S.}~\bibnamefont {Xu}}, \bibinfo {author} {\bibfnamefont
  {P.}~\bibnamefont {Cao}}, \bibinfo {author} {\bibfnamefont {Y.}~\bibnamefont
  {Zhou}}, \bibinfo {author} {\bibfnamefont {W.}~\bibnamefont {Cao}},\ and\
  \bibinfo {author} {\bibfnamefont {P.}~\bibnamefont {Lu}},\ }\bibfield
  {title} {\bibinfo {title} {{Identification of tunneling and multiphoton
  ionization in intermediate Keldysh parameter regime}},\ }\href
  {http://www.opticsexpress.org/abstract.cfm?URI=oe-27-5-6471} {\bibfield
  {journal} {\bibinfo  {journal} {Opt. Express}\ }\textbf {\bibinfo {volume}
  {27}},\ \bibinfo {pages} {6471} (\bibinfo {year} {2019})}\BibitemShut
  {NoStop}%
\bibitem [{\citenamefont {Yuan}(2019)}]{Yuan_2019}%
  \BibitemOpen
  \bibfield  {author} {\bibinfo {author} {\bibfnamefont {M.}~\bibnamefont
  {Yuan}},\ }\bibfield  {title} {\bibinfo {title} {{Direct probing of tunneling
  time in strong-field ionization processes by time-dependent wave packets}},\
  }\href@noop {} {\bibfield  {journal} {\bibinfo  {journal} {Opt. Express}\
  }\textbf {\bibinfo {volume} {27}},\ \bibinfo {pages} {6502} (\bibinfo {year}
  {2019})}\BibitemShut {NoStop}%
\bibitem [{\citenamefont {Kheifets}(2020)}]{Kheifets_2020}%
  \BibitemOpen
  \bibfield  {author} {\bibinfo {author} {\bibfnamefont {A.~S.}\ \bibnamefont
  {Kheifets}},\ }\bibfield  {title} {\bibinfo {title} {{The attoclock and the
  tunneling time debate}},\ }\href@noop {} {\bibfield  {journal} {\bibinfo
  {journal} {J. Phys. B}\ }\textbf {\bibinfo {volume} {53}},\ \bibinfo {pages}
  {72001} (\bibinfo {year} {2020})}\BibitemShut {NoStop}%
\bibitem [{\citenamefont {Yusofsani}\ and\ \citenamefont
  {Kolesik}(2020)}]{Kolesik_2020}%
  \BibitemOpen
  \bibfield  {author} {\bibinfo {author} {\bibfnamefont {S.}~\bibnamefont
  {Yusofsani}}\ and\ \bibinfo {author} {\bibfnamefont {M.}~\bibnamefont
  {Kolesik}},\ }\bibfield  {title} {\bibinfo {title} {Quantum tunneling time:
  Insights from an exactly solvable model},\ }\href@noop {} {\bibfield
  {journal} {\bibinfo  {journal} {Phys. Rev. A}\ }\textbf {\bibinfo {volume}
  {101}},\ \bibinfo {pages} {052121} (\bibinfo {year} {2020})}\BibitemShut
  {NoStop}%
\bibitem [{\citenamefont {Yu}\ \emph {et~al.}(2022)\citenamefont {Yu},
  \citenamefont {Liu}, \citenamefont {Li}, \citenamefont {Yan}, \citenamefont
  {Cao}, \citenamefont {Tan}, \citenamefont {Liang}, \citenamefont {Guo},
  \citenamefont {Cao}, \citenamefont {Lan}, \citenamefont {Zhang},
  \citenamefont {Zhou},\ and\ \citenamefont {Lu}}]{Yu_2022}%
  \BibitemOpen
  \bibfield  {author} {\bibinfo {author} {\bibfnamefont {M.}~\bibnamefont
  {Yu}}, \bibinfo {author} {\bibfnamefont {K.}~\bibnamefont {Liu}}, \bibinfo
  {author} {\bibfnamefont {M.}~\bibnamefont {Li}}, \bibinfo {author}
  {\bibfnamefont {J.}~\bibnamefont {Yan}}, \bibinfo {author} {\bibfnamefont
  {C.}~\bibnamefont {Cao}}, \bibinfo {author} {\bibfnamefont {J.}~\bibnamefont
  {Tan}}, \bibinfo {author} {\bibfnamefont {J.}~\bibnamefont {Liang}}, \bibinfo
  {author} {\bibfnamefont {K.}~\bibnamefont {Guo}}, \bibinfo {author}
  {\bibfnamefont {W.}~\bibnamefont {Cao}}, \bibinfo {author} {\bibfnamefont
  {P.}~\bibnamefont {Lan}}, \bibinfo {author} {\bibfnamefont {Q.}~\bibnamefont
  {Zhang}}, \bibinfo {author} {\bibfnamefont {Y.}~\bibnamefont {Zhou}},\ and\
  \bibinfo {author} {\bibfnamefont {P.}~\bibnamefont {Lu}},\ }\bibfield
  {title} {\bibinfo {title} {{Full experimental determination of tunneling time
  with attosecond-scale streaking method}},\ }\href
  {https://www.nature.com/articles/s41377-022-00911-8} {\bibfield  {journal}
  {\bibinfo  {journal} {Light: Science \& Applications}\ }\textbf {\bibinfo
  {volume} {11}},\ \bibinfo {pages} {1} (\bibinfo {year} {2022})}\BibitemShut
  {NoStop}%
\bibitem [{\citenamefont {Klaiber}\ \emph {et~al.}(2022)\citenamefont
  {Klaiber}, \citenamefont {Lv}, \citenamefont {Sukiasyan}, \citenamefont
  {Bakucz~Can\'ario}, \citenamefont {Hatsagortsyan},\ and\ \citenamefont
  {Keitel}}]{Klaiber_2022}%
  \BibitemOpen
  \bibfield  {author} {\bibinfo {author} {\bibfnamefont {M.}~\bibnamefont
  {Klaiber}}, \bibinfo {author} {\bibfnamefont {Q.~Z.}\ \bibnamefont {Lv}},
  \bibinfo {author} {\bibfnamefont {S.}~\bibnamefont {Sukiasyan}}, \bibinfo
  {author} {\bibfnamefont {D.}~\bibnamefont {Bakucz~Can\'ario}}, \bibinfo
  {author} {\bibfnamefont {K.~Z.}\ \bibnamefont {Hatsagortsyan}},\ and\
  \bibinfo {author} {\bibfnamefont {C.~H.}\ \bibnamefont {Keitel}},\ }\bibfield
   {title} {\bibinfo {title} {Reconciling conflicting approaches for the
  tunneling time delay in strong field ionization},\ }\href
  {https://doi.org/10.1103/PhysRevLett.129.203201} {\bibfield  {journal}
  {\bibinfo  {journal} {Phys. Rev. Lett.}\ }\textbf {\bibinfo {volume} {129}},\
  \bibinfo {pages} {203201} (\bibinfo {year} {2022})}\BibitemShut {NoStop}%
\bibitem [{\citenamefont {MacColl}(1932)}]{MacColl_1932}%
  \BibitemOpen
  \bibfield  {author} {\bibinfo {author} {\bibfnamefont {L.~A.}\ \bibnamefont
  {MacColl}},\ }\bibfield  {title} {\bibinfo {title} {{Note on the Transmission
  and Reflection of Wave Packets by Potential Barriers}},\ }\href@noop {}
  {\bibfield  {journal} {\bibinfo  {journal} {Phys. Rev.}\ }\textbf {\bibinfo
  {volume} {40}},\ \bibinfo {pages} {621} (\bibinfo {year} {1932})}\BibitemShut
  {NoStop}%
\bibitem [{\citenamefont {Eisenbud}(1948)}]{Eisenbud}%
  \BibitemOpen
  \bibfield  {author} {\bibinfo {author} {\bibfnamefont {L.}~\bibnamefont
  {Eisenbud}},\ }\emph {\bibinfo {title} {{The Formal Properties of Nuclear
  Collisions.}}},\ \href@noop {} {Ph.D. thesis},\ \bibinfo  {school} {Princeton
  University} (\bibinfo {year} {1948})\BibitemShut {NoStop}%
\bibitem [{\citenamefont {Wigner}\ and\ \citenamefont
  {Eisenbud}(1947)}]{EisenbudWigner}%
  \BibitemOpen
  \bibfield  {author} {\bibinfo {author} {\bibfnamefont {E.~P.}\ \bibnamefont
  {Wigner}}\ and\ \bibinfo {author} {\bibfnamefont {L.}~\bibnamefont
  {Eisenbud}},\ }\bibfield  {title} {\bibinfo {title} {{Higher Angular Momenta
  and Long Range Interaction in Resonance Reactions}},\ }\href@noop {}
  {\bibfield  {journal} {\bibinfo  {journal} {Phys. Rev.}\ }\textbf {\bibinfo
  {volume} {72}},\ \bibinfo {pages} {29} (\bibinfo {year} {1947})}\BibitemShut
  {NoStop}%
\bibitem [{\citenamefont {Wigner}(1955)}]{wigner_1955}%
  \BibitemOpen
  \bibfield  {author} {\bibinfo {author} {\bibfnamefont {E.~P.}\ \bibnamefont
  {Wigner}},\ }\bibfield  {title} {\bibinfo {title} {Lower limit for the energy
  derivative of the scattering phase shift},\ }\href@noop {} {\bibfield
  {journal} {\bibinfo  {journal} {Phys. Rev.}\ }\textbf {\bibinfo {volume}
  {98}},\ \bibinfo {pages} {145} (\bibinfo {year} {1955})}\BibitemShut
  {NoStop}%
\bibitem [{\citenamefont {Smith}(1960)}]{Smith_1960}%
  \BibitemOpen
  \bibfield  {author} {\bibinfo {author} {\bibfnamefont {F.~T.}\ \bibnamefont
  {Smith}},\ }\bibfield  {title} {\bibinfo {title} {Lifetime matrix in
  collision theory},\ }\href {https://doi.org/10.1103/PhysRev.118.349}
  {\bibfield  {journal} {\bibinfo  {journal} {Phys. Rev.}\ }\textbf {\bibinfo
  {volume} {118}},\ \bibinfo {pages} {349} (\bibinfo {year}
  {1960})}\BibitemShut {NoStop}%
\bibitem [{\citenamefont {Baz'}(1966)}]{Baz_1966}%
  \BibitemOpen
  \bibfield  {author} {\bibinfo {author} {\bibfnamefont {A.~I.}\ \bibnamefont
  {Baz'}},\ }\bibfield  {title} {\bibinfo {title} {{Lifetime of intermediate
  states}},\ }\href@noop {} {\bibfield  {journal} {\bibinfo  {journal} {Yad.
  Fiz.}\ }\textbf {\bibinfo {volume} {4}} (\bibinfo {year} {1966})}\BibitemShut
  {NoStop}%
\bibitem [{\citenamefont {B\"uttiker}\ and\ \citenamefont
  {Landauer}(1982)}]{Butttiker_1982}%
  \BibitemOpen
  \bibfield  {author} {\bibinfo {author} {\bibfnamefont {M.}~\bibnamefont
  {B\"uttiker}}\ and\ \bibinfo {author} {\bibfnamefont {R.}~\bibnamefont
  {Landauer}},\ }\bibfield  {title} {\bibinfo {title} {Traversal time for
  tunneling},\ }\href {https://doi.org/10.1103/PhysRevLett.49.1739} {\bibfield
  {journal} {\bibinfo  {journal} {Phys. Rev. Lett.}\ }\textbf {\bibinfo
  {volume} {49}},\ \bibinfo {pages} {1739} (\bibinfo {year}
  {1982})}\BibitemShut {NoStop}%
\bibitem [{\citenamefont {Pollak}\ and\ \citenamefont
  {Miller}(1984)}]{Pollak_1984}%
  \BibitemOpen
  \bibfield  {author} {\bibinfo {author} {\bibfnamefont {E.}~\bibnamefont
  {Pollak}}\ and\ \bibinfo {author} {\bibfnamefont {W.~H.}\ \bibnamefont
  {Miller}},\ }\bibfield  {title} {\bibinfo {title} {New physical
  interpretation for time in scattering theory},\ }\href
  {https://doi.org/10.1103/PhysRevLett.53.115} {\bibfield  {journal} {\bibinfo
  {journal} {Phys. Rev. Lett.}\ }\textbf {\bibinfo {volume} {53}},\ \bibinfo
  {pages} {115} (\bibinfo {year} {1984})}\BibitemShut {NoStop}%
\bibitem [{\citenamefont {Steinberg}(1995)}]{Steinberg_1995}%
  \BibitemOpen
  \bibfield  {author} {\bibinfo {author} {\bibfnamefont {A.~M.}\ \bibnamefont
  {Steinberg}},\ }\bibfield  {title} {\bibinfo {title} {How much time does a
  tunneling particle spend in the barrier region?},\ }\href
  {https://doi.org/10.1103/PhysRevLett.74.2405} {\bibfield  {journal} {\bibinfo
   {journal} {Phys. Rev. Lett.}\ }\textbf {\bibinfo {volume} {74}},\ \bibinfo
  {pages} {2405} (\bibinfo {year} {1995})}\BibitemShut {NoStop}%
\bibitem [{\citenamefont {Czirj\'ak}\ \emph {et~al.}(2000)\citenamefont
  {Czirj\'ak}, \citenamefont {Kopold}, \citenamefont {Becker}, \citenamefont
  {Kleber},\ and\ \citenamefont {Schleich}}]{Czirjak_2000}%
  \BibitemOpen
  \bibfield  {author} {\bibinfo {author} {\bibfnamefont {A.}~\bibnamefont
  {Czirj\'ak}}, \bibinfo {author} {\bibfnamefont {R.}~\bibnamefont {Kopold}},
  \bibinfo {author} {\bibfnamefont {W.}~\bibnamefont {Becker}}, \bibinfo
  {author} {\bibfnamefont {M.}~\bibnamefont {Kleber}},\ and\ \bibinfo {author}
  {\bibfnamefont {W.}~\bibnamefont {Schleich}},\ }\bibfield  {title} {\bibinfo
  {title} {The wigner function for tunneling in a uniform static electric
  field1dedicated to marlan o. scully on the occasion of his 60th birthday.1},\
  }\href {https://doi.org/https://doi.org/10.1016/S0030-4018(99)00591-X}
  {\bibfield  {journal} {\bibinfo  {journal} {Optics Communications}\ }\textbf
  {\bibinfo {volume} {179}},\ \bibinfo {pages} {29} (\bibinfo {year}
  {2000})}\BibitemShut {NoStop}%
\bibitem [{\citenamefont {Peres}(1980)}]{Peres_1980}%
  \BibitemOpen
  \bibfield  {author} {\bibinfo {author} {\bibfnamefont {A.}~\bibnamefont
  {Peres}},\ }\bibfield  {title} {\bibinfo {title} {Measurement of time by
  quantum clocks},\ }\href@noop {} {\bibfield  {journal} {\bibinfo  {journal}
  {Am. J. Phys.}\ }\textbf {\bibinfo {volume} {48}},\ \bibinfo {pages} {552}
  (\bibinfo {year} {1980})}\BibitemShut {NoStop}%
\bibitem [{\citenamefont {Landauer}\ and\ \citenamefont
  {Martin}(1994)}]{Landauer_1994}%
  \BibitemOpen
  \bibfield  {author} {\bibinfo {author} {\bibfnamefont {R.}~\bibnamefont
  {Landauer}}\ and\ \bibinfo {author} {\bibfnamefont {T.}~\bibnamefont
  {Martin}},\ }\bibfield  {title} {\bibinfo {title} {{Barrier interaction time
  in tunneling}},\ }\href@noop {} {\bibfield  {journal} {\bibinfo  {journal}
  {Rev. Mod. Phys.}\ }\textbf {\bibinfo {volume} {66}},\ \bibinfo {pages} {217}
  (\bibinfo {year} {1994})}\BibitemShut {NoStop}%
\bibitem [{\citenamefont {Hauge}\ and\ \citenamefont
  {St\o{}vneng}(1989)}]{Hauge_1998}%
  \BibitemOpen
  \bibfield  {author} {\bibinfo {author} {\bibfnamefont {E.~H.}\ \bibnamefont
  {Hauge}}\ and\ \bibinfo {author} {\bibfnamefont {J.~A.}\ \bibnamefont
  {St\o{}vneng}},\ }\bibfield  {title} {\bibinfo {title} {Tunneling times: a
  critical review},\ }\href {https://doi.org/10.1103/RevModPhys.61.917}
  {\bibfield  {journal} {\bibinfo  {journal} {Rev. Mod. Phys.}\ }\textbf
  {\bibinfo {volume} {61}},\ \bibinfo {pages} {917} (\bibinfo {year}
  {1989})}\BibitemShut {NoStop}%
\bibitem [{\citenamefont {Muga}\ \emph {et~al.}(2002)\citenamefont {Muga},
  \citenamefont {Egusquiza},\ and\ \citenamefont {Mayato}}]{Muga}%
  \BibitemOpen
  \bibinfo {editor} {\bibfnamefont {J.~G.}\ \bibnamefont {Muga}}, \bibinfo
  {editor} {\bibfnamefont {I.~L.}\ \bibnamefont {Egusquiza}},\ and\ \bibinfo
  {editor} {\bibfnamefont {R.~S.}\ \bibnamefont {Mayato}},\ eds.,\ \href@noop
  {} {\emph {\bibinfo {title} {{Time in Quantum Mechanics}}}},\ \bibinfo
  {series} {Lecture Notes in Physics}\ No.~\bibinfo {number} {72}\ (\bibinfo
  {publisher} {Springer},\ \bibinfo {year} {2002})\BibitemShut {NoStop}%
\bibitem [{\citenamefont {de~Carvalho}\ and\ \citenamefont
  {Nussenzveig}(2002)}]{deCarvalho}%
  \BibitemOpen
  \bibfield  {author} {\bibinfo {author} {\bibfnamefont {C.~A.~A.}\
  \bibnamefont {de~Carvalho}}\ and\ \bibinfo {author} {\bibfnamefont {H.~M.}\
  \bibnamefont {Nussenzveig}},\ }\bibfield  {title} {\bibinfo {title} {{Time
  delay}},\ }\href@noop {} {\bibfield  {journal} {\bibinfo  {journal} {Phys.
  Rep.}\ }\textbf {\bibinfo {volume} {364}},\ \bibinfo {pages} {83} (\bibinfo
  {year} {2002})}\BibitemShut {NoStop}%
\bibitem [{\citenamefont {Davies}(2005)}]{Davies_2005}%
  \BibitemOpen
  \bibfield  {author} {\bibinfo {author} {\bibfnamefont {P.~C.~W.}\
  \bibnamefont {Davies}},\ }\bibfield  {title} {\bibinfo {title} {Quantum
  tunneling time},\ }\href@noop {} {\bibfield  {journal} {\bibinfo  {journal}
  {Am. J. Phys.}\ }\textbf {\bibinfo {volume} {73}},\ \bibinfo {pages} {23}
  (\bibinfo {year} {2005})}\BibitemShut {NoStop}%
\bibitem [{\citenamefont {Ramos}\ \emph {et~al.}(2020)\citenamefont {Ramos},
  \citenamefont {Spierings},\ and\ \citenamefont {Racicot}}]{Ramos_2020}%
  \BibitemOpen
  \bibfield  {author} {\bibinfo {author} {\bibfnamefont {R.}~\bibnamefont
  {Ramos}}, \bibinfo {author} {\bibfnamefont {D.}~\bibnamefont {Spierings}},\
  and\ \bibinfo {author} {\bibfnamefont {I.~t.}\ \bibnamefont {Racicot}},\
  }\bibfield  {title} {\bibinfo {title} {{Measurement of the time spent by a
  tunnelling atom within the barrier region}},\ }\href@noop {} {\bibfield
  {journal} {\bibinfo  {journal} {Nature}\ }\textbf {\bibinfo {volume} {583}},\
  \bibinfo {pages} {529–532} (\bibinfo {year} {2020})}\BibitemShut {NoStop}%
\bibitem [{\citenamefont {Feuerstein}\ and\ \citenamefont
  {Thumm}(2003)}]{Feuerstein_2003}%
  \BibitemOpen
  \bibfield  {author} {\bibinfo {author} {\bibfnamefont {B.}~\bibnamefont
  {Feuerstein}}\ and\ \bibinfo {author} {\bibfnamefont {U.}~\bibnamefont
  {Thumm}},\ }\bibfield  {title} {\bibinfo {title} {{On the computation of
  momentum distributions within wavepacket propagation calculations}},\
  }\href@noop {} {\bibfield  {journal} {\bibinfo  {journal} {J. Phys. B}\
  }\textbf {\bibinfo {volume} {36}},\ \bibinfo {pages} {707} (\bibinfo {year}
  {2003})}\BibitemShut {NoStop}%
\bibitem [{\citenamefont {Wang}\ \emph {et~al.}(2013)\citenamefont {Wang},
  \citenamefont {Tian},\ and\ \citenamefont {Eberly}}]{Wang_2013}%
  \BibitemOpen
  \bibfield  {author} {\bibinfo {author} {\bibfnamefont {X.}~\bibnamefont
  {Wang}}, \bibinfo {author} {\bibfnamefont {J.}~\bibnamefont {Tian}},\ and\
  \bibinfo {author} {\bibfnamefont {J.~H.}\ \bibnamefont {Eberly}},\ }\bibfield
   {title} {\bibinfo {title} {{Extended Virtual Detector Theory for
  Strong-Field Atomic Ionization}},\ }\href@noop {} {\bibfield  {journal}
  {\bibinfo  {journal} {Phys. Rev. Lett.}\ }\textbf {\bibinfo {volume} {110}},\
  \bibinfo {pages} {243001} (\bibinfo {year} {2013})}\BibitemShut {NoStop}%
\bibitem [{\citenamefont {Becker}\ \emph {et~al.}(2002)\citenamefont {Becker},
  \citenamefont {Grasbon}, \citenamefont {Kopold}, \citenamefont
  {Milo\u{s}evi{\'{c}}}, \citenamefont {Paulus},\ and\ \citenamefont
  {Walther}}]{Becker_2002}%
  \BibitemOpen
  \bibfield  {author} {\bibinfo {author} {\bibfnamefont {W.}~\bibnamefont
  {Becker}}, \bibinfo {author} {\bibfnamefont {F.}~\bibnamefont {Grasbon}},
  \bibinfo {author} {\bibfnamefont {R.}~\bibnamefont {Kopold}}, \bibinfo
  {author} {\bibfnamefont {D.~B.}\ \bibnamefont {Milo\u{s}evi{\'{c}}}},
  \bibinfo {author} {\bibfnamefont {G.~G.}\ \bibnamefont {Paulus}},\ and\
  \bibinfo {author} {\bibfnamefont {H.}~\bibnamefont {Walther}},\ }\bibfield
  {title} {\bibinfo {title} {{Above-threshold ionization: from classical
  features to quantum effects}},\ }\href@noop {} {\bibfield  {journal}
  {\bibinfo  {journal} {Adv. Atom. Mol. Opt. Phys.}\ }\textbf {\bibinfo
  {volume} {48}},\ \bibinfo {pages} {35} (\bibinfo {year} {2002})}\BibitemShut
  {NoStop}%
\bibitem [{\citenamefont {Can\'ario}\ \emph {et~al.}(2021)\citenamefont
  {Can\'ario}, \citenamefont {Klaiber}, \citenamefont {Hatsagortsyan},\ and\
  \citenamefont {Keitel}}]{Canario_2021}%
  \BibitemOpen
  \bibfield  {author} {\bibinfo {author} {\bibfnamefont {D.~B.}\ \bibnamefont
  {Can\'ario}}, \bibinfo {author} {\bibfnamefont {M.}~\bibnamefont {Klaiber}},
  \bibinfo {author} {\bibfnamefont {K.~Z.}\ \bibnamefont {Hatsagortsyan}},\
  and\ \bibinfo {author} {\bibfnamefont {C.~H.}\ \bibnamefont {Keitel}},\
  }\bibfield  {title} {\bibinfo {title} {Role of reflections in the generation
  of a time delay in strong-field ionization},\ }\href
  {https://doi.org/10.1103/PhysRevA.104.033103} {\bibfield  {journal} {\bibinfo
   {journal} {Phys. Rev. A}\ }\textbf {\bibinfo {volume} {104}},\ \bibinfo
  {pages} {033103} (\bibinfo {year} {2021})}\BibitemShut {NoStop}%
\bibitem [{\citenamefont {Keldysh}(1964)}]{Keldysh_1965}%
  \BibitemOpen
  \bibfield  {author} {\bibinfo {author} {\bibfnamefont {L.~V.}\ \bibnamefont
  {Keldysh}},\ }\bibfield  {title} {\bibinfo {title} {{Ionization in the field
  of a strong electromagnetic wave}},\ }\href@noop {} {\bibfield  {journal}
  {\bibinfo  {journal} {Zh. Eksp. Teor. Fiz.}\ }\textbf {\bibinfo {volume}
  {47}},\ \bibinfo {pages} {1945} (\bibinfo {year} {1964})}\BibitemShut
  {NoStop}%
\bibitem [{\citenamefont {Faisal}(1973)}]{Faisal_1973}%
  \BibitemOpen
  \bibfield  {author} {\bibinfo {author} {\bibfnamefont {F.~H.~M.}\
  \bibnamefont {Faisal}},\ }\bibfield  {title} {\bibinfo {title} {Multiple
  absorption of laser photons by atoms},\ }\href@noop {} {\bibfield  {journal}
  {\bibinfo  {journal} {J. Phys. B}\ }\textbf {\bibinfo {volume} {6}},\
  \bibinfo {pages} {L89} (\bibinfo {year} {1973})}\BibitemShut {NoStop}%
\bibitem [{\citenamefont {Reiss}(1980)}]{Reiss_1980}%
  \BibitemOpen
  \bibfield  {author} {\bibinfo {author} {\bibfnamefont {H.~R.}\ \bibnamefont
  {Reiss}},\ }\bibfield  {title} {\bibinfo {title} {Effect of an intesne
  electromagnetic field on a weakly bound system},\ }\href@noop {} {\bibfield
  {journal} {\bibinfo  {journal} {Phys. Rev. A}\ }\textbf {\bibinfo {volume}
  {22}},\ \bibinfo {pages} {1786} (\bibinfo {year} {1980})}\BibitemShut
  {NoStop}%
\bibitem [{\citenamefont {Wolkow}(1935)}]{Volkov_1935}%
  \BibitemOpen
  \bibfield  {author} {\bibinfo {author} {\bibfnamefont {D.~M.}\ \bibnamefont
  {Wolkow}},\ }\bibfield  {title} {\bibinfo {title} {{{\"{U}}ber eine Klasse
  von L{\"{o}}sungen der Diracschen Gleichung}},\ }\href@noop {} {\bibfield
  {journal} {\bibinfo  {journal} {Z. Phys.}\ }\textbf {\bibinfo {volume}
  {94}},\ \bibinfo {pages} {250} (\bibinfo {year} {1935})}\BibitemShut
  {NoStop}%
\bibitem [{\citenamefont {Milo\ifmmode \check{s}\else
  \v{s}\fi{}evi\ifmmode~\acute{c}\else \'{c}\fi{}}(2014)}]{Milosevic_2014}%
  \BibitemOpen
  \bibfield  {author} {\bibinfo {author} {\bibfnamefont {D.~B.}\ \bibnamefont
  {Milo\ifmmode \check{s}\else \v{s}\fi{}evi\ifmmode~\acute{c}\else
  \'{c}\fi{}}},\ }\bibfield  {title} {\bibinfo {title} {Forward- and
  backward-scattering quantum orbits in above-threshold ionization},\ }\href
  {https://doi.org/10.1103/PhysRevA.90.063414} {\bibfield  {journal} {\bibinfo
  {journal} {Phys. Rev. A}\ }\textbf {\bibinfo {volume} {90}},\ \bibinfo
  {pages} {063414} (\bibinfo {year} {2014})}\BibitemShut {NoStop}%
\bibitem [{\citenamefont {Milo\ifmmode \check{s}\else
  \v{s}\fi{}evi\ifmmode~\acute{c}\else \'{c}\fi{}}\ and\ \citenamefont
  {Becker}(2022)}]{Milosevic_2022}%
  \BibitemOpen
  \bibfield  {author} {\bibinfo {author} {\bibfnamefont {D.~B.}\ \bibnamefont
  {Milo\ifmmode \check{s}\else \v{s}\fi{}evi\ifmmode~\acute{c}\else
  \'{c}\fi{}}}\ and\ \bibinfo {author} {\bibfnamefont {W.}~\bibnamefont
  {Becker}},\ }\bibfield  {title} {\bibinfo {title} {Negative-travel-time
  quantum orbits in strong-field ionization by an elliptically polarized laser
  field},\ }\href {https://doi.org/10.1103/PhysRevA.105.L031103} {\bibfield
  {journal} {\bibinfo  {journal} {Phys. Rev. A}\ }\textbf {\bibinfo {volume}
  {105}},\ \bibinfo {pages} {L031103} (\bibinfo {year} {2022})}\BibitemShut
  {NoStop}%
\bibitem [{\citenamefont {\ifmmode \check{C}\else
  \v{C}\fi{}erki\ifmmode~\acute{c}\else \'{c}\fi{}}\ \emph
  {et~al.}(2009)\citenamefont {\ifmmode \check{C}\else
  \v{C}\fi{}erki\ifmmode~\acute{c}\else \'{c}\fi{}}, \citenamefont
  {Hasovi\ifmmode~\acute{c}\else \'{c}\fi{}}, \citenamefont {Milo\ifmmode
  \check{s}\else \v{s}\fi{}evi\ifmmode~\acute{c}\else \'{c}\fi{}},\ and\
  \citenamefont {Becker}}]{Cerkic_2009}%
  \BibitemOpen
  \bibfield  {author} {\bibinfo {author} {\bibfnamefont {A.}~\bibnamefont
  {\ifmmode \check{C}\else \v{C}\fi{}erki\ifmmode~\acute{c}\else \'{c}\fi{}}},
  \bibinfo {author} {\bibfnamefont {E.}~\bibnamefont
  {Hasovi\ifmmode~\acute{c}\else \'{c}\fi{}}}, \bibinfo {author} {\bibfnamefont
  {D.~B.}\ \bibnamefont {Milo\ifmmode \check{s}\else
  \v{s}\fi{}evi\ifmmode~\acute{c}\else \'{c}\fi{}}},\ and\ \bibinfo {author}
  {\bibfnamefont {W.}~\bibnamefont {Becker}},\ }\bibfield  {title} {\bibinfo
  {title} {{High-order above-threshold ionization beyond the first-order Born
  approximation}},\ }\href@noop {} {\bibfield  {journal} {\bibinfo  {journal}
  {Phys. Rev. A}\ }\textbf {\bibinfo {volume} {79}},\ \bibinfo {pages} {033413}
  (\bibinfo {year} {2009})}\BibitemShut {NoStop}%
\bibitem [{\citenamefont {Milo\v{s}evi\'c}(2014)}]{Milosevic_2014a}%
  \BibitemOpen
  \bibfield  {author} {\bibinfo {author} {\bibfnamefont {D.~B.}\ \bibnamefont
  {Milo\v{s}evi\'c}},\ }\bibfield  {title} {\bibinfo {title} {{Low-frequency
  approximation for above-threshold ionization by a laser pulse: Low-energy
  forward rescattering}},\ }\href@noop {} {\bibfield  {journal} {\bibinfo
  {journal} {Phys. Rev. A}\ }\textbf {\bibinfo {volume} {90}},\ \bibinfo
  {pages} {063423} (\bibinfo {year} {2014})}\BibitemShut {NoStop}%
\bibitem [{\citenamefont {Klaiber}\ \emph {et~al.}(2013)\citenamefont
  {Klaiber}, \citenamefont {Yakaboylu}, \citenamefont {Bauke}, \citenamefont
  {Hatsagortsyan},\ and\ \citenamefont {Keitel}}]{Klaiber_2013}%
  \BibitemOpen
  \bibfield  {author} {\bibinfo {author} {\bibfnamefont {M.}~\bibnamefont
  {Klaiber}}, \bibinfo {author} {\bibfnamefont {E.}~\bibnamefont {Yakaboylu}},
  \bibinfo {author} {\bibfnamefont {H.}~\bibnamefont {Bauke}}, \bibinfo
  {author} {\bibfnamefont {K.~Z.}\ \bibnamefont {Hatsagortsyan}},\ and\
  \bibinfo {author} {\bibfnamefont {C.~H.}\ \bibnamefont {Keitel}},\ }\bibfield
   {title} {\bibinfo {title} {Under-the-barrier dynamics in laser-induced
  relativistic tunneling},\ }\href
  {https://doi.org/10.1103/PhysRevLett.110.153004} {\bibfield  {journal}
  {\bibinfo  {journal} {Phys. Rev. Lett.}\ }\textbf {\bibinfo {volume} {110}},\
  \bibinfo {pages} {153004} (\bibinfo {year} {2013})}\BibitemShut {NoStop}%
\bibitem [{\citenamefont {Lohr}\ \emph {et~al.}(1997)\citenamefont {Lohr},
  \citenamefont {Kleber}, \citenamefont {Kopold},\ and\ \citenamefont
  {Becker}}]{Lohr_1997}%
  \BibitemOpen
  \bibfield  {author} {\bibinfo {author} {\bibfnamefont {A.}~\bibnamefont
  {Lohr}}, \bibinfo {author} {\bibfnamefont {M.}~\bibnamefont {Kleber}},
  \bibinfo {author} {\bibfnamefont {R.}~\bibnamefont {Kopold}},\ and\ \bibinfo
  {author} {\bibfnamefont {W.}~\bibnamefont {Becker}},\ }\bibfield  {title}
  {\bibinfo {title} {Above-threshold ionization in the tunneling regime},\
  }\href {https://doi.org/10.1103/PhysRevA.55.R4003} {\bibfield  {journal}
  {\bibinfo  {journal} {Phys. Rev. A}\ }\textbf {\bibinfo {volume} {55}},\
  \bibinfo {pages} {R4003} (\bibinfo {year} {1997})}\BibitemShut {NoStop}%
\bibitem [{\citenamefont {Krajewska}\ \emph {et~al.}(2010)\citenamefont
  {Krajewska}, \citenamefont {Kami{\'{n}}ski},\ and\ \citenamefont
  {W{\'{o}}dkiewicz}}]{Krajewska_2010}%
  \BibitemOpen
  \bibfield  {author} {\bibinfo {author} {\bibfnamefont {K.}~\bibnamefont
  {Krajewska}}, \bibinfo {author} {\bibfnamefont {J.~Z.}\ \bibnamefont
  {Kami{\'{n}}ski}},\ and\ \bibinfo {author} {\bibfnamefont {K.}~\bibnamefont
  {W{\'{o}}dkiewicz}},\ }\bibfield  {title} {\bibinfo {title} {{Zero-range
  interaction in arbitrary dimensions and in the presence of external
  forces}},\ }\href@noop {} {\bibfield  {journal} {\bibinfo  {journal} {Opt.
  Commun.}\ }\textbf {\bibinfo {volume} {283}},\ \bibinfo {pages} {843}
  (\bibinfo {year} {2010})}\BibitemShut {NoStop}%
\bibitem [{\citenamefont {Su\'arez}\ \emph {et~al.}(2015)\citenamefont
  {Su\'arez}, \citenamefont {Chac\'on}, \citenamefont {Ciappina}, \citenamefont
  {Biegert},\ and\ \citenamefont {Lewenstein}}]{Suarez_2015}%
  \BibitemOpen
  \bibfield  {author} {\bibinfo {author} {\bibfnamefont {N.}~\bibnamefont
  {Su\'arez}}, \bibinfo {author} {\bibfnamefont {A.}~\bibnamefont {Chac\'on}},
  \bibinfo {author} {\bibfnamefont {M.~F.}\ \bibnamefont {Ciappina}}, \bibinfo
  {author} {\bibfnamefont {J.}~\bibnamefont {Biegert}},\ and\ \bibinfo {author}
  {\bibfnamefont {M.}~\bibnamefont {Lewenstein}},\ }\bibfield  {title}
  {\bibinfo {title} {Above-threshold ionization and photoelectron spectra in
  atomic systems driven by strong laser fields},\ }\href
  {https://doi.org/10.1103/PhysRevA.92.063421} {\bibfield  {journal} {\bibinfo
  {journal} {Phys. Rev. A}\ }\textbf {\bibinfo {volume} {92}},\ \bibinfo
  {pages} {063421} (\bibinfo {year} {2015})}\BibitemShut {NoStop}%
\end{thebibliography}%

 \end{document}